\documentclass[12pt, draftclsnofoot, onecolumn]{IEEEtran}
\usepackage{cite}
\usepackage{array}
\usepackage{framed}
\usepackage{xcolor}
\usepackage{bm}
\usepackage{booktabs}	

\newenvironment{eqn}%
{\begin{equation}\begin{aligned}}
{\end{aligned}\end{equation}}

{\begin{small}\begin{eqnarray}\begin{aligned}}%
{\end{aligned}\end{eqnarray}\end{small}}

\newcommand{\norm}[1]{\Vert #1 \Vert}

\newcommand{\multiline}[1]{%
  \begin{tabularx}{\dimexpr\linewidth-\ALG@thistlm}[t]{@{}X@{}}
    #1
  \end{tabularx}
}

\ifCLASSOPTIONcompsoc
  \usepackage[caption=false,font=normalsize,labelfont=sf,textfont=sf]{subfig}
\else
  \usepackage[caption=false,font=footnotesize]{subfig}
\fi

\usepackage{dblfloatfix}
\ifCLASSINFOpdf
   \usepackage[pdftex]{graphicx}
   \DeclareGraphicsExtensions{.pdf,.jpeg,.png}
\else
\fi
\usepackage{amsmath}
\usepackage{amssymb}
\usepackage{amsthm}     
\usepackage{blkarray}   
\usepackage{color}  
\usepackage{url}

\usepackage{algorithm, tabularx} 
\usepackage[noend]{algpseudocode} 

\allowdisplaybreaks
\mathchardef\mhyphen="2D 
\begin{document}
\title{\huge Deep-Learning Based Auction-Driven Beamforming for Wireless Information and Power Transfer}
\author{\IEEEauthorblockN{Ali Bayat,~\IEEEmembership{Student Member,~IEEE,} and Sonia A{\"\i}ssa,~\IEEEmembership{Fellow,~IEEE}}
\thanks{The authors are with the Institut National de la Recherche Scientifique (INRS), Montreal, QC, Canada (e-mail: ali.bayat@inrs.ca; sonia.aissa@inrs.ca).}
\thanks{This work was supported by a Discovery Grant from the Natural Sciences and Engineering Research Council (NSERC) of Canada.}
}


\maketitle


\begin{abstract}
\noindent In this paper, we design a deep learning based resource allocation framework, in the form of an auction, for simultaneous information and power transfer from a hybrid access point (AP) to information devices and energy harvesting devices, respectively. Using Myerson's lemma and the concept of virtual welfare maximization, we develop an optimal dominant-strategy incentive-compatible mechanism for the AP to maximize its expected revenue, based on the devices' bid profiles, valuation distributions, demand profiles, and channel state information. In so doing, we formulate the revenue maximization problem, which is a mixed-integer non-linear program, and propose an efficient Branch-and-Bound (BnB) algorithm to solve the problem using semidefinite relaxation technique in each branch. Since the problem has exponential time complexity, using BnB algorithms can be impractical for real-time applications. To circumvent this, a deep neural network (DNN) is proposed, and trained to predict the optimal mechanism for beamforming the data and the energy towards the information and energy devices, respectively. We use the BnB algorithm to solve the problem offline and populate the training dataset. The proposed DNN architecture is indeed a multi-layer perceptron, which is trained well to map the heterogeneous input to the desired output with high accuracy. Furthermore, we propose a heuristic iterative solution whose accuracy performance is comparable to that of the DNN-based solution. The heuristic solution has polynomial time complexity whereas the DNN-based solution has linear time complexity.
\end{abstract}
\begin{IEEEkeywords}
\noindent Auction theory; Beamforming; Deep learning; Simultaneous wireless information and power transfer (SWIPT).
\end{IEEEkeywords}
\IEEEdisplaynontitleabstractindextext
%
%
\section{Introduction}
\subsection{Context and Motivation}
In the era of massive machine-type communications and the Internet of Things, simultaneous wireless information and power transfer (SWIPT) is arising as a chief technology for supplying energy to, and information exchange with, numerous low-power devices in a plethora of applications. By judiciously superposing information and power transfer, SWIPT is not only envisioned to help in improving system performance and service quality, e.g., in terms of spectral efficiency, time delay, and energy efficiency, but also in managing interferences, thereby increasing substantiality \cite{Krikidis2014SWIPT}.

In spite of many theoretical advances for implementing SWIPT systems (cf. \cite{Clerckx2019WIPTFundaments} and references therein), in order to take benefit of SWIPT technology and materialize its large-scale penetration in wireless systems and networks, several challenges remain to be addressed. A chief challenge is the design of beamforming-based resource allocation algorithms of low computational complexity and suitable for real-time applications. One such complexity happens when looking for the optimal set of user equipments (UEs) in the network,\footnote{The terms user, UE, and device, will be used interchangeably throughout the paper.} including information receivers (IRs) and energy receivers (ERs), to maximize a specific objective function, e.g., revenue of the service provider or social welfare of the users. In these instances, the optimization problem generally lies in the class of mixed-integer non-linear programming (MINLP) problems.

The main issue is how to manage and allocate the limited network resources in an efficient way. Resource management/allocation in wireless networks are envisioned to be mainly based on user-centric online service provisioning schemes, where users act as economical agents that compete for their target services according to competition market models such as \textit{auctions} \cite{Zhang2013AuctionRA, Tadayon2018Auction}. Auctions are probably the best marketing model when entities tend to behave selfishly to maximize their own utility. The inclusion of the economical behavior of the agents in managing the resources grants the agents more degrees of freedom to bid for their desired service based on their demands and their monetary budgets. It is foreseen that marketing and pricing tasks will no more be settled statistically in an offline service-level agreement, but rather be of dynamic, online, and cross-layered design nature~\cite{Niyato2009CRAuction,Zhang2012AuctionRACR,Habiba2018AuctionNV5G,Tadayon2018Auction}.

In this context, there is a large body of work on the application of game theory and auction theory for the design and analysis of resource allocation mechanisms in wireless communication networks~\cite{Zhang2013AuctionRA, Habiba2018AuctionNV5G, Tadayon2018Auction}. In particular, a good survey on auction approaches for resource allocation in wireless systems is carried out in \cite{Zhang2013AuctionRA}. In \cite{Habiba2018AuctionNV5G}, the authors conducted a contemporary survey on the application of auction mechanisms for the virtualization of wireless networks, in particular 5G cellular networks. When it comes to the application of game (auction) theory in wireless powered communications, the works in \cite{Niyato2014WPTGT, Chen2015WPCNGT, Sarma2016EHGT, Zheng2017WPRNGT, Bayat2017WPCN, Bayat2018Critical, Zina-WCNC2021, Saif-GC2021} are important to recall. In the said works, a wireless power transfer system (WPT) in which a central unit transmits energy to ERs via energy beamforming is considered. So far, and to the best of our knowledge, leveraging the merits of auctions for the design of SWIPT systems remains unexplored.

As aforementioned, one major challenge in designing dynamic auction mechanisms for wireless systems is the computational complexity that may hinder real-time implementation. As an example, when the mechanism objective is to maximize the revenue of the service provider, e.g., the AP in a SWIPT network, finding the optimal set of users that satisfies this objective is usually formulated as a MINLP problem, which is in general NP-hard with no efficient global optimal algorithm available.
The optimal solutions to MINLP problems are usually achieved by the Branch-and-Bound (BnB) algorithms. However, the computational complexity of such algorithms is exponential, making them impractical for real-time implementation. Sub-optimal solutions, such as heuristic algorithms, can certainly reduce the computational complexity, however, they usually suffer two problems: i) the difficulty in controlling the performance gap between the sub-optimal solution and the optimal one, and ii) although they converge faster than the BnB algorithms, most of them are still too computationally complex to be suitable for real-time implementation.

One promising solution approach to tackle these computational complexities is by applying machine learning (ML) techniques, in particular deep learning (DL), which is considered as the key branch of ML, with focus on learning and prediction by means of training (deep) neural networks (DNNs). In \cite{Dutting2019AuctionDlrn}, the authors modeled multi-item auctions as a multi-layer neural network and framed an optimal auction design as a constrained learning problem. The proposed DNN, called RegretNet, works for bidders with discrete additive, unit-demand, and combinatorial valuations. As such, it is not applicable for bidders with continuous demand and bid values. Also, RegretNet is not directly applicable for time-varying environments like in the case of wireless networks, where the availability of the commodities depend hugely on the channel conditions. In \cite{Lee2020RADNN}, a DL-based resource allocation scheme for device-to-device communications was proposed. Therein, DL was leveraged for learning efficient pruning policies in BnB algorithms aimed for solving the corresponding MINLP problem. DL was also leveraged in \cite{Lee2020RADNN} to find the optimal branching of the corresponding BnB algorithm in an iterative algorithm. It is noted that the latter work does not address computational analysis. In \cite{Dong2020DLrnRA5G}, a resource allocation algorithm based on deep transfer learning was proposed for ultra-reliable low-latency communications.

Though there are works such as \cite{Varasteh2019LrnSWIPT} and \cite{Chun2018SWIPTEHDLrn} that applied DL and reinforcement learning techniques in SWIPT systems, none is related to finding the optimal resource allocation solution while tackling the computational complexity issue. For instance, in \cite{Varasteh2019LrnSWIPT}, the authors aimed for the optimal modulations for point-to-point SWIPT system via the application of NN-based auto-encoders; \cite{Chun2018SWIPTEHDLrn} proposed an adaptive rate and energy harvesting interval control algorithm based on the model-free reinforcement learning technique for SWIPT; and \cite{AlEryani2020EHDRLrn} applied deep reinforcement learning for simultaneous energy harvesting and information transmission in a MIMO (multi-input multi-output) full-duplex system.
\subsection{Contributions}
Focused on SWIPT, in this work we consider a single-band heterogeneous SWIPT network, in which a multi-antenna AP attempts to maximize its revenue by beamforming the information and energy towards the single-antenna IRs and ERs, respectively. The network operates in an auction framework. Based on the bids, requested service levels, learned valuation distributions, and CSI of the users, the AP---as the auctioneer and the seller---aims to find the optimal set of users and the optimal pricing that maximize its revenue while encouraging the users to bid truthfully.

We formulate this mechanism design problem as a MINLP problem, and solve it by using a proposed efficient BnB algorithm while applying semidefinite programming (SDP) technique in each branch. Since solving this problem via the BnB algorithm is time-greedy, particularly when the number of devices in the network is fairly large, the branching algorithms are no more practical due to their excessive computation delays---exponential time complexity. We tackle the time complexity problem by applying DL, particularly DNNs, to output the solution almost in real time. Before that, we first propose a heuristic iterative algorithm based on \textit{goodness factors} which are separately defined for the two types of users, i.e.,  IRs and ERs. In spite of the good accuracy of the heuristic algorithm, we reach for the more computational-efficient DL-based solutions. For such, we first interpret the problem of finding the optimal set of users as a multi-label classification ML problem. Then, we propose a DNN-based architecture---multi-layer perceptron (MLP) to be more specific---to solve the classification problem and find the desired optimal set of users, after which the optimal beamforming vectors can be found by one-time running of a semidefinite relaxation (SDR)-based algorithm or a faster uplink-downlink duality (UDD)-based one \cite{Schubert2004BeamformSINRconstraints, Xu2014MuMISOSWIPT}.

To gather the required training data, we solve the MINLP problem offline for a very large number of realizations of the heterogeneous input data to the AP, save the training dataset and the obtained training labels in a database, and then use the latter to train the proposed DNN so as to estimate the allocation rule, i.e., the optimal set of users, of the proposed revenue-maximizing auction mechanism. In stack contrast to most DL-based approaches in the field of wireless communications, where the input data is homogeneous, e.g., the channel matrix, here we face a heterogeneous input data which makes the decision-making on the best DNN architecture and input pre-processing layers very challenging.

After trying several network architecture candidates, including convolutional neural networks, residual neural networks, and mixed-input neural networks, we found the MLP architecture with the goodness factor-based preprocessing layer to be the most promising. We analytically compare the computational complexity of the global optimal BnB algorithm (with the inner branch algorithms being SDR or UDD) to that of the heuristic algorithm and the DNN-based solution. In particular, we show that the heuristic algorithm follows the accuracy performance of the DNN-based solution but with polynomial time complexity compared to the linear complexity of the DNN approach.\footnote{While this work was in progress, preliminary results related to the special case where the network is comprised of information devices only and, specifically, on the application of DL for the data beamforming to IRs with the objective of maximizing the social welfare of the users, were submitted to IEEE Globecom 2020 \cite{Bayat2020BFDlrnAuction}. The present paper has the following key differences with the latter: i) designing revenue-maximizing mechanism for a SWIPT network whereas \cite{Bayat2020BFDlrnAuction} proposes a DNN for social welfare maximization in a wireless information transfer (WIT) network, ii) here, an efficient BnB algorithm to find the optimal solutions is proposed, and iii) a heuristic goodness-based algorithm is also proposed and its performance is compared for different numbers of served IR and ER devices.}
\subsection{Organization}
In detailing the aforementioned contributions, the remainder of the paper is organized as follows. Section \ref{sec:SystemModel} models the SWIPT network, both physically and economically. In Section \ref{sec:ProblemFormulation}, we formulate the revenue maximization problem. In Section \ref{sec:FeasibleSet}, we find the feasible allocation set as part of the solution of the revenue-maximization problem using SDR-based technique. Therein, we also investigate the two particular cases of the SWIPT model: WIT and WPT. In section \ref{sec:rev_max_mech}, the optimal revenue-maximizing mechanism is obtained. Section \ref{sec:Algos} presents an efficient BnB algorithm to obtain the optimal allocation rule along with a heuristic iterative algorithm which delivers a sub-optimal solution. Section \ref{sec:NN} presents the proposed DNN-based solution, and discusses its architecture and training. Therein, a computational complexity analysis is also provided. In section \ref{sec:PerformanceEvaluation}, we evaluate the performance of the proposed DNN for three network operation modes: SWIPT, WIT, and WPT. We also compare the accuracy of the proposed DNN-based and heuristic methods. Finally, Section \ref{sec:Conclusion} concludes the paper.
\subsubsection*{Notations} The following set of notations will be adopted throughout the paper. Vectors and matrices are written with bold lower- and upper-case letters, respectively. Symbols $(.)^T$ and $(.)^H$ denote the transpose and conjugate transpose operators. The identity matrix of order $m$ is denoted by $\mathbf{I}_m$, tr($\mathbf{A}$) is the trace of square matrix $A$, and $\mathbf{0}$ is a zero vector with proper dimension. The $l_1$-norm and $l_2$-norm (Euclidean) are denoted by $\norm{.}_{1}$ and $\norm{.}$, respectively, and $\mathbb{E}[.]$ stands for mathematical expectation.

\section{The SWIPT Network Modeling}\label{sec:SystemModel}
Modeling the SWIPT network consists of the physical modeling of the network elements, the economical (bidding) behavioral modeling of the devices, and the framework of the auction.
\subsection{Physical Modeling}
The SWIPT network consists of a AP serving multiple devices within a shared spectrum band. The AP is equipped with $M$ antennas, and has a power budget of $P$ Watts. A total number of $K=I+J$ UEs coexist in the network, with $I$ denoting the number of IRs, and $J$ being the number of ERs. Set $\mathcal{I}=\{1,\dots,I\}$ contains the indices of the information devices, and $\mathcal{J}=\{1,\dots,J\}$ is the index set of the energy devices. All devices are equipped with single antennas, and work in half-duplex mode similar to the AP. The AP is to provide wireless energy to ERs, and to send data to IRs. Without loss of generality, we consider linear precoding at the AP, such that each ER/IR is assigned with one dedicated energy/information transmission beam.\footnote{Later in Section \ref{sec:FeasibleSet}, we will show that only one beamforing vector is sufficient for all ERs.} The signal transmitted from the AP is given by
\begin{equation}
\mathbf{x} = \sum_{  i\in \mathcal{I}} \mathbf{w}_i s_{\mathrm{IR}_i} + \sum_{  j\in \mathcal{J}} \mathbf{v}_j s_{\mathrm{ER}_j},
\end{equation}
where, for any $\mathrm{IR}_{i}, i\in {\cal I}$, $\mathbf{w}_i\in \mathbb{C}^{M\times 1}$ is the beamforming vector and $s_{\mathrm{IR}_i}$ is the information-bearing signal, and where, for any $\mathrm{ER}_j, j\in \cal J$, $\mathbf{v}_j\in \mathbb{C}^{M\times 1}$ is the beamforming vector and $s_{\mathrm{ER}_j}$ is the energy-carrying signal.

For the information signals, we assume Gaussian inputs, i.e., the $s_{\mathrm{IR}_i}$'s are i.i.d. circularly-symmetric complex Gaussian (CSCG) random variables with zero mean and unit variance, denoted by $s_{\mathrm{IR}_i} \sim {\cal CN}(0,1),   i\in \mathcal{I}$. For the energy signals, since $s_{\mathrm{ER}_j},   j\in \mathcal{J}$, carries no information, it can be any arbitrary signal that satisfies the radio regulations on microwave radiation \cite{Xia2015WPT}. Without loss of generality, we assume that the $s_{\mathrm{ER}_j}$'s are independent white sequences from an arbitrary distribution with $\mathbb{E}\left [\vert s_{\mathrm{ER}_j}\vert^{2}\right]=1,  j\in \mathcal{J}$. Given the limit $P$ on the AP's transmit power, the constraint $\mathbb{E}\left [\mathbf{x}^{H} \mathbf{x}\right ]= \sum_{i\in \mathcal{I}}\Vert \mathbf{w}_i \Vert^{2} +\sum_{j\in \mathcal{J}}\Vert \mathbf{v}_j\Vert ^{2} \leq P$ must hold.

The fading channels between the transmitter and the receivers are quasi-static, i.e., channel coefficients are assumed to be fixed during the channel coherence time. Denote $\mathbf{h}_{i}=(h_{i,1},\dots,h_{i,M})^{T}$ and $\mathbf{g}_{j}=(g_{j,1},\dots,g_{j,M})^{T}$ as the channel vectors from the AP to $\mathrm{IR}_i$ and $\mathrm{ER}_j$, respectively, where $\norm{\mathbf{h}_i}^2 = \sigma_{\mathbf{h}_i}^2$ and $\norm{\mathbf{g}_j}^2 = \sigma_{\mathbf{g}_j}^2$ for $i\in\cal I$ and $j\in\cal J$. The channel vectors $\mathbf{h}_i, i\in \mathcal{I}$, and $\mathbf{g}_j, j\in \mathcal{J}$, are drawn independently from continuous distribution functions $F_{h_i}(\mathbf{h}_i)$ and $F_{g_j}(\mathbf{g}_j)$, with $h_{i,m}$ and $g_{j,m}$ being the complex channel gains from the $m^{\rm th}$ antenna of the AP array, $m\in\{1,\dots,M\}$, to $\mathrm{IR}_i$ and $\mathrm{ER}_j$, respectively. These channel vectors are assumed to be perfectly tracked at the devices and fed back to the AP via an error-free zero-delay feedback channel.

The received base-band equivalent signal at $\mathrm{IR}_i, i\in \cal I$, is
\begin{equation}
y_{i} = \mathbf{h}_i^T \mathbf{x} + z_{i},
\end{equation}
where $z_{i}\sim {\cal CN}(0,\sigma _{i}^{2})$ is the i.i.d. Gaussian noise. Therefore, the signal-to-interference-plus-noise ratio (SINR) of information receiver $\mathrm{IR}_i, i\in \cal I$, can be written as
\begin{eqn}\label{eq:Gammak}
{\Gamma}_i & = \frac{|\mathbf{h}_{i}^{T}\mathbf{w}_i|^2}{\sum_{k\in\mathcal{I}, k\neq i} |\mathbf{h}_{i}^{T}\mathbf{w}_k|^2 + \sum_{j\in\cal J} |\mathbf{h}_{i}^{T}\mathbf{v}_j|^2 + \sigma_i^2	} \\ &= \frac{\mathbf{w}_i^{H}\mathbf{R}_{i}\mathbf{w}_i}{\sum_{k\in\mathcal{I}, k\neq i} \mathbf{w}_k^{H}\mathbf{R}_{i}\mathbf{w}_k + \sum_{j\in\cal J} \mathbf{v}_j^{H}\mathbf{R}_{i}\mathbf{v}_j + \sigma_i^2},
\end{eqn}%
where $\mathbf{R}_{i} = \mathbb{E}[\mathbf{h}_i\mathbf{h}_i^H]$ is the covariance matrix, which for the case of full CSI knowledge becomes $\mathbf{R}_{i} = \mathbf{h}_i\mathbf{h}_i^H$.

The SINR is directly related to the device's performance indicators such as the bit error rate (BER) and the data rate. For example, under
a fixed BER and assuming quadrature-amplitude modulation, a practical achievable rate can be computed as $R_i=\log (1 + \Gamma_i / \Upsilon)\,\rm bps/Hz$, in which $\Upsilon$ denotes the SNR (signal-to-noise ratio) gap to capacity. The SNR gap is always greater than 1 (0 dB), and it gives an approximate relation between the SINR and the rate.

The received power at ER$_j$, $j\in \cal J$, is given by
\begin{eqn}\label{eq:Qk}
Q_{j}&=\sum_{i\in\mathcal{I}}\vert \mathbf{g}_{j}^{T}\mathbf{w}_i\vert ^{2} + \sum_{k\in\mathcal{J}}\vert \mathbf{g}_{j}^{T}\mathbf{v}_k\vert ^{2}\\&= \sum_{i\in\mathcal{I}} \mathbf{w}_i^{H}\mathbf{C}_{j}\mathbf{w}_i + \sum_{k\in\mathcal{J}} \mathbf{v}_k^{H}\mathbf{C}_{j}\mathbf{v}_k,
\end{eqn}%
where $\mathbf{C}_{j} = \mathbb{E}[\mathbf{g}_j\mathbf{g}_j^H]$ is the covariance matrix, which for the case of full CSI knowledge becomes $\mathbf{C}_{j} = \mathbf{g}_j\mathbf{g}_j^H$.
\subsection{Bidding Modeling}
The AP as the service seller plays the role of the auctioneer as well. The devices play the roles of bidders which have different service valuations sending their bids in each round of auction to get served by the AP. The AP solicits the devices' bids in a sealed fashion, i.e., the devices are not aware of each others' bids.

It is assumed that all devices have non-zero service requests and play in all auction rounds. Each $\mathrm{IR}_i$, $i\in\cal I$, sends its service request in the form of the minimum SINR, $\gamma_i$, required to receive its data in the auction duration $\tau_a$, by bidding $b_{\mathrm{IR}_i}$. Each $\mathrm{ER}_j$, $j\in\cal J$, requests $q_j$ units of energy for the auction duration $\tau_a$ by bidding $b_{\mathrm{ER}_j}$. The AP knows in advance that $\mathrm{IR}_i$ and $\mathrm{ER}_j$ draw their private valuations $\nu_{\mathrm{IR}_i}$ and $\nu_{\mathrm{ER}_j}$ from the distributions $\mathcal{F}_{\mathrm{IR}_i}$ and $\mathcal{F}_{\mathrm{ER}_j}$ in each round of auction.\footnote{By private values, it is meant that the values are unknown to the seller (AP) and to other bidders. Valuation is the maximum willingness-to-pay of an agent for the commodity being sold. Also, the distributions $\mathcal{F}_{\mathrm{IR}_i}$ and $\mathcal{F}_{\mathrm{ER}_j}$ can be estimated (learned) from the UEs' bids history in the past auctions \cite{Nisan2007AlgorithmicGameTheory_Ebook, Jiang2005OnlineAuctions}.} These distributions are assumed independent, but not necessarily identical. Without loss of generality, we set $\tau_a = 1$. Also, let $\bm{\gamma}=(\gamma_1,\dots,\gamma_{I})^{T}$ and $\mathbf{q}=(q_1,\dots,q_J)^{T}$ denote the demand profiles of IRs and ERs, respectively, and denote $\mathbf{b}_{\rm IR}=(b_{\mathrm{IR}_1},\dots, b_{\mathrm{IR}_I})^{T}$ and $\mathbf{b}_{\rm ER}=(b_{\mathrm{ER}_1},\dots, b_{\mathrm{ER}_J})^{T}$ as the bid profiles of IRs and ERs, respectively.
\subsection{Auction Framework}\label{sec:Timeframe}
\begin{figure}[h]
\centering
\includegraphics[width=0.6\linewidth]{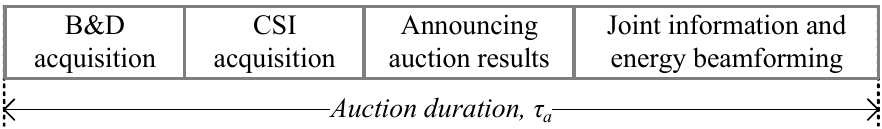}
\caption{Breakout of each auction round.}
\label{fig:AuctionEpisodes}
\end{figure}

We consider a single-parameter (or single-dimensional) auction environment, where the outcomes of the AP as the auction mechanism designer are two rules: (i) the {\it allocation rule} $\mathbf{a}=(\substack{\mathbf{a}_{\rm IR}\\ \mathbf{a}_{\rm ER}})$, where $\mathbf{a}_{\rm IR}=(a_{\mathrm{IR}_1},\dots,a_{\mathrm{IR}_I})^{T}$ and $\mathbf{a}_{\rm ER}=(a_{\mathrm{ER}_1},\dots,a_{\mathrm{ER}_J})^{T}$, with each element of vector $\mathbf{a}_{\rm IR}$ ($\mathbf{a}_{\rm ER}$) being an indicator for whether $\mathrm{IR}_i$ ($\mathrm{ER}_j$) is to be served or not and, thus, $\mathbf{a}_{\rm IR}\in \{0, 1\}^{I}$ ($\mathbf{a}_{\rm ER}\in \{0, 1\}^{J}$); and (ii) the {\it payment rule} $\mathbf{p}=(\substack{\mathbf{p}_{\rm IR}\\ \mathbf{p}_{\rm ER}})$, where $\mathbf{p}_{\rm IR}=(p_{\mathrm{IR}_1},\dots,p_{\mathrm{IR}_I})^{T}$ and $\mathbf{p}_{\rm ER}=(p_{\mathrm{ER}_1},\dots,p_{\mathrm{ER}_J})^{T}$, with each element of vector $\mathbf{p}_{\rm IR}$ ($\mathbf{p}_{\rm ER}$) denoting the amount that $\mathrm{IR}_i$ ($\mathrm{ER}_j$) is required to pay the auctioneer, i.e., the AP, during any round of auction. As depicted in Fig. \ref{fig:AuctionEpisodes}, each round of auction is composed of four parts: (i) bids-and-demands (B\&D) acquisition, (ii) CSI acquisition, (iii) auction results announcing, and (iv) beamforming.

To find the optimal beamforming vectors $\mathbf{w}_{i}^{\star}$, $i\in\cal I$, and $\mathbf{v}_{j}^{\star}$, $j\in\cal J$, the key step is to find the optimal allocation vector $\mathbf{a}^{\star}$ and the optimal payment vector $\mathbf{p}^{\star}$ that maximize the revenue of the AP. Then, the AP sends the pairs $(a_{\mathrm{IR}_i}^{\star}, p_{\mathrm{IR}_i}^{\star})$ and $(a_{\mathrm{ER}_j}^{\star}, p_{\mathrm{ER}_j}^{\star})$ to ${\rm IR}_i$ and ${\rm ER}_j$, respectively, as the auction result to let them know whether they have won the auction and how much they should pay in case of a win.\footnote{Allocation (payment) ``rule" and allocation (payment) ``vector" are used interchangeably throughout this paper.}

At the beginning of each auction round, the AP solicits the UEs for their demands and their corresponding bids during the B\&D-acquisition period. Then, the AP acquires the CSI.\footnote{To estimate the CSI, one can consider either \textit{one-way training} by assuming channel reciprocity, or \textit{two-way training} which requires each receiver to perform channel estimation followed by feedback to the AP, which in turn will consume additional energy. In practice, there exists a design tradeoff especially for the ERs: higher accuracy for both channel estimation and feedback reporting may lead to higher harvested energy due to the transmit beamforming gain, but also induces higher energy consumption that can even offset the harvested energy gain \cite{Xu2014EB1bit}. For simplicity, in this work we assume that such energy consumption at ERs is negligible compared to their harvested energy, and that the training time is also negligible compared to the auction period. It is assumed that the auction duration is equal to the channel coherence time, and is much larger that the CSI acquisition duration.} Afterwards, the AP solves for the optimal allocation rule and payment rule that maximize its expected revenue while keeping the devices incentivized to play truthfully, announces the allocation-payment rules to all UEs to let them know who are going to get service and how much to pay for it and, finally, performs joint beamforming to transfer the information and power to the chosen UEs.
\subsection{Utility Functions}
First off, for ease of writing, let us define $\mathcal{K}=\{1,\dots,I-1,I,I+1,\dots,I+J\}$ as the ordered set of all the UEs in the SWIPT network. With this notational convention, $\mathrm{UE}_k = \mathrm{IR}_k$ for $k\in \cal I$ and $\mathrm{UE}_k = \mathrm{ER}_{k-I}$ when $k\in\{I+1,\dots, I+J\}$. We also denote the bid vector by $\mathbf{b}=(\substack{\mathbf{b}_{\rm IR}\\ \mathbf{b}_{\rm ER}})$, and the demand vector by $\mathbf{d}=(\substack{\bm{\gamma}\\\mathbf{q}})$.

In designing optimal mechanisms for auctions, there are two fundamental objectives: \textit{social welfare}, a.k.a. social surplus, and \textit{revenue}, a.k.a. profit.

Revenue generated from the allocation-payment pair ($\mathbf{a},\mathbf{p}$) constitutes the utility of the AP. Specifically, it is the cumulative payment of the bidders minus the service cost $c$:\footnote{Note that the service cost $c$ depends on the allocation rule $\mathbf{a}$.}
\begin{equation}\label{eq:uAP1}
u_{\rm AP} = \sum_{k\in \mathcal{K}} p_k - c.
\end{equation}
For $\mathrm{UE}_k, k\in\cal{K}$, with valuation $\nu_k$\footnote{This parameter should not be mistaken with the energy beamforming vector $\mathbf{v}_j$ of ${\rm ER}_j$.} , the utility---assuming the quasi-linear model \cite{Hartline2013MechanismDesign_Ebook}---is defined as follows:\footnote{In quasi-linear utility model, an agent goal is to choose a bid that maximizes the difference between his valuation and his payment.}
\begin{equation}
u_k = \nu_k\, a_k - p_k.
\end{equation}
It should be emphasized that $a_k$ and $p_k$ depend on the bid profile $\mathbf{b}$.

The social welfare, resulting from the allocation rule $\mathbf{a}$, is the cumulative valuations of all agents in the auction minus the service cost, i.e.,
\begin{eqn}\label{eq:Svx}
S(\bm{\nu}, \mathbf{a})= \sum_{k\in \mathcal{K}} \nu_k\,a_k - c=\mathbf{a}^{T}\bm{\nu} - c,
\end{eqn}
where vector $\bm{\nu}=(\substack{\bm{\nu}_{\rm IR}\\ \bm{\nu}_{\rm ER}})$, in which $\bm{\nu}_{\rm IR}=(\nu_{\mathrm{IR}_1},\dots,\nu_{\mathrm{IR}_I})^{T}$ and $\bm{\nu}_{\rm ER}=(\nu_{\mathrm{ER}_1},\dots,\nu_{\mathrm{ER}_J})^{T}$, holds the valuations of the UEs. The CSI acquisition costs can be regarded as part of the total service cost $c$. In this work, just for simplicity and without affecting the contributions of the paper, the cost function is assumed zero. Thus, $u_{\rm AP} = \sum_{k\in\cal K} p_k= \sum_{i\in\cal I}p_{\mathrm{IR}_i} + \sum_{j\in\cal J}p_{\mathrm{ER}_j}$, and $S(\bm{\nu}, \mathbf{a}) = \mathbf{a}^{T}\bm{\nu} = {\mathbf{a}_{\rm IR}}^{T}\bm{\nu}_{\rm IR} + {\mathbf{a}_{\rm ER}}^{T}\bm{\nu}_{\rm ER}$. Bid $b_k$ is the number that $\mathrm{UE}_k$, $k\in\cal K$, declares to the AP as payment for its demand $d_k$, whereas valuation $\nu_k$ is the true belief of the device about its demand $d_k$.
\section{The Problem Formulation}\label{sec:ProblemFormulation}
Finding the optimal revenue maximizing (Rmax) mechanism is equivalent to finding the optimal allocation and payment rules. These rules can be obtained by solving the following Rmax problem:
\begin{equation}\label{eq:optim2}
\underset{\mathbf{a}\in\mathcal{A}_\mathrm{F}}{\rm max} ~ \mathbb{E}_{\bm{\nu}}\left [ u_{\rm AP}\right ]= \mathbb{E}_{\bm{\nu}}\left [ \sum_{k\in\mathcal{K}} p_k \right],
\end{equation}%
where the expectation is w.r.t. the distribution $\mathcal{F}_{\cal K}=\prod_{k\in \cal K}\mathcal{F}_{k}=\prod_{i\in \cal I} \mathcal{F}_{\mathrm{IR}_i} \prod_{j\in \cal J} \mathcal{F}_{\mathrm{ER}_j}$ over the bidders' valuations $\nu_k, k\in\cal K$. In (\ref{eq:optim2}), $\mathcal{A}_{\rm F}$ is the set of all feasible allocation vectors, which depends on the CSI of the channels $\mathbf{h}_i$, $i\in \cal I$, and $\mathbf{g}_j$, $j\in\cal J$, and on the devices' demand profile $\mathbf{d}$. An allocation vector is deemed feasible if the minimum power required to satisfy the demand constraints of the subset of devices represented by that allocation vector is less than the power budget of the AP.

The problem in (\ref{eq:optim2}) states that the optimal allocation vector $\mathbf{a}^{\star}$ corresponds to the feasible set of users that results in the
largest sum-payment. However, the payment rule $\textbf{p}$ itself should be carefully found to keep the devices incentivized to play truthfully. In fact, there are three unknowns in (\ref{eq:optim2}): $\mathbf{a}^{\star}$, $\mathbf{p}^{\star}$ and $\mathcal{A}_{\rm F}$. The latter is independent of the first two unknowns and is what is firstly obtained in Section \ref{sec:FeasibleSet}. The revenue-maximizing mechanism which constitutes the optimal pair $(\mathbf{a}^{\star}, \mathbf{p}^{\star})$ will be obtained in Section \ref{sec:rev_max_mech}.
\section{Finding the Feasible Allocation Set $\mathcal{A}_{\rm F}$}\label{sec:FeasibleSet}
In order for the AP to find the feasible allocation vectors out of all possible realizations of allocation vectors $\mathbf{a}^{(l)}\in\{0,1\}^{K}$, $l\in\{0,1,\dots, 2^{K}-1\}$, a series of non-convex optimization problems should be solved. Let us correspond a subset $A^{(l)}\subset \cal K$ to each $\mathbf{a}^{(l)}=({a}_1^{(l)},\dots,{a}_K^{(l)})^{T}$ for any {$l\in\{0,1,\dots, 2^{K}-1\}$}, so that $\mathrm{UE}_k\in A^{(l)}$ {\it iff} ${a}_k^{(l)} = 1$, $k \in \cal K$. Superscript $l$ is the decimal representation of the binary vector $\mathbf{a}^{(l)}$. We put the IR-type and the ER-type devices in $A^{(l)}$ into the subsets ${A}_{\rm IR}^{(l)}$ and ${A}_{\rm ER}^{(l)}$, respectively. Similarly, $\mathbf{a}^{(l)}=\left (\substack{\mathbf{a}_{\rm IR}^{(l)}\\\mathbf{a}_{\rm ER}^{(l)}}\right )$.

In general, we face a MINLP problem. In the worst-case, there would be $2^K - 1$ optimization problems to solve in order to find $\mathcal{A}_{\rm F}$. Hence, the problem has exponential time complexity.

To illustrate how $\mathbf{a}^{(l)}$ is interpreted, consider the example of $I=3$ and $J=1$, which results in $K=4$ devices and the index set $\mathcal{K}=\{1,2,3,4\}$. Then, $\mathbf{a}^{(0)} = (0,0,0,0)^T$ corresponds to $A^{(0)}=\emptyset$, meaning that no devices are chosen for allocation; $\mathbf{a}^{(1)} = (0,0,0,1)^T$ corresponds to $A^{(1)}=\{4\}$, meaning that no IR devices are chosen and the only ER device is selected for allocation; and $\mathbf{a}^{(2)} = (0,0,1,0)^T$ corresponds to $A^{(2)}=\{3\}$ which means that only $\mathrm{IR}_{3}$ is chosen for allocation.

Each problem corresponding to $\mathbf{a}^{(l)}$ is a power minimization problem, which consists of finding the minimum transmit power required by the AP to fulfill the demands of the chosen UEs in subset $A^{(l)}$, and is formulated as follows
\begin{eqn}\label{eq:qcqp1}
P_{A^{(l)}}^{\rm min}=& \underset{\substack{\mathbf{w}_i,\,i\in A_{\rm IR}^{(l)}\\\mathbf{v}_j,\,j\in A_{\rm ER}^{(l)}}}{\rm min} \sum_{i\in A_{\rm IR}^{(l)}} \Vert \mathbf{w}_i \Vert^2 + \sum_{j\in A_{\rm ER}^{(l)}} \Vert \mathbf{v}_j \Vert^2,\\
{\rm s.t.}:~&\Gamma_i \geq \gamma_i,   i\in A_{\rm IR}^{(l)}\\
&Q_j \geq q_j,   j\in A_{\rm ER}^{(l)},
\end{eqn}
where $\Gamma_i$ and $Q_j$ are shown in (\ref{eq:Gammak}) and (\ref{eq:Qk}), respectively. We recall that the $\gamma$-parameters imply the SINRs required for achieving certain data rates at the IRs, and that the $q$-parameters describe the amount of input power needed by ERs to meet an equivalent output DC power requirement. If the found $P_{A^{(l)}}^{\min}$ in (\ref{eq:qcqp1}) is within the AP's power budget $P$, then the corresponding $A^{(l)}$ is a feasible subset of UEs and, equivalently,
$\mathbf{a}^{(l)}\in \mathcal{A}_{\rm F}$. Furthermore, the feasible subsets are \textit{downward-closed}, which means that subsets of feasible sets are feasible themselves.

By defining $\mathbf{W}_i=\mathbf{w}_i\mathbf{w}_i^{H}$, $i\in\mathcal{I}$, and $\mathbf{V}_j=\mathbf{v}_j\mathbf{v}_j^{H}$, $j\in\mathcal{J}$, we can write the original quadratically-constraint quadratic problem (QCQP) (\ref{eq:qcqp1}) in the following format which can be efficiently solved by SDR techniques.
%
\begin{eqn}\label{eq:socp1}
P_{A^{(l)}}^{\rm min}=& \underset{\substack{\mathbf{W}_i,\,i\in A_{\rm IR}^{(l)}\\\mathbf{V}_j,\,j\in A_{\rm ER}^{(l)}}}{\rm min} \sum_{i\in A_{\rm IR}^{(l)}}{\rm tr}(\mathbf{W}_i) + \sum_{j\in A_{\rm ER}^{(l)}}{\rm tr}(\mathbf{V}_j)\\
{\rm s.t.}: ~ &{\rm tr}(\mathbf{R}_{i} \mathbf{W}_i) - \gamma_i \Big(\sum_{\substack{k\in {A_{\mathrm{IR}}}^{(l)}\\k\neq i}}{\rm tr}(\mathbf{R}_{i} \mathbf{W}_{k}) + \sum_{j\in {A_{\mathrm{ER}}^{(l)}}} {\rm tr}(\mathbf{R}_{i} \mathbf{V}_{j})\Big) \geq \gamma_i \sigma_i^2,~i\in A_{\rm IR}^{(l)},\\
&\sum_{k\in A_{\rm IR}^{(l)}}\mathrm{tr}(\mathbf{C}_{j} \mathbf{W}_{k}) + \sum_{k\in A_{\rm ER}^{(l)}}\mathrm{tr}(\mathbf{C}_{j} \mathbf{V}_{k})  \geq q_j,~j\in A_{\rm ER}^{(l)},\\
& \mathbf{W}_{i} \succeq 0, ~i\in A_{\rm IR}^{(l)},\\
& \mathbf{V}_{j} \succeq 0, ~j\in A_{\rm ER}^{(l)},
\end{eqn}%
%
where $\mathbf{R}_{i} \succeq 0$ and $\mathbf{C}_{i} \succeq 0$. The optimization problem in (\ref{eq:socp1}) is of conic form, and can be solved using standard tools such as CVX \cite{Grant2014CVX}. It can be understood from \cite{Xu2014MuMISOSWIPT} that the SDR problem in (\ref{eq:socp1}) is tight meaning that the solutions $\mathbf{W}_i^{\star}$'s and $\mathbf{V}_j^{\star}$'s are rank-$1$ maximum. This will be further explained in subsection \ref{ssec:Back2IER}.

Next, to further discuss the solution to (\ref{eq:socp1}), we study two particular models of the network: (i) the WIT scenario, where only IR devices are active, and (ii) the WPT scenario, which corresponds to the network where only the ER devices are active. After solving the problem related to either scenario with conventional optimization tools and techniques, we argue the inherent time-greediness feature of these techniques for real-time applications like our network model, which makes such techniques loose their applicability. We propose alternative sub-optimal solutions in subsection \ref{ssec:goodness} for each scenario.
\subsection{Wireless Information Transfer Network}\label{ssec:WIT}
Considering IR devices only in the network, i.e., $\cal K = I$, problem (\ref{eq:qcqp1}) becomes
\begin{eqn}\label{eq:qcqp2}
P_{A^{(l)}}^{\rm min}&= \underset{\mathbf{w}_i,\,i\in A^{(l)}}{\rm min} \sum_{i\in A^{(l)}} \Vert \mathbf{w}_i \Vert^2\\
{\rm s.t.}:&~\Gamma_i \geq \gamma_i, i\in A^{(l)}.
\end{eqn}%
Two well-known methods can be used to solve the sub-problems in (\ref{eq:qcqp2}): semidefinite relaxation based technique \cite{Bengtsson1999BFSDR}, and uplink-downlink duality based algorithm \cite{Schubert2004BeamformSINRconstraints}.
\subsubsection{SDR-Based Solution}
Recalling $\mathbf{W}_i=\mathbf{w}_i\mathbf{w}_i^{H},   i\in\mathcal{K}$, and relaxing the rank-1 constraint $\mathrm{rank}(\mathbf{W}_i)=1,   i\in\cal{K}$, we can write the second-order cone program (SOCP) shown in (\ref{eq:qcqp2}) in the semidefinite relaxed program format shown in (\ref{eq:socpUEs}). The resulting problem can be solved by standard tools such as CVX \cite{Grant2014CVX}.
\begin{eqn}\label{eq:socpUEs}
P_{A^{(l)}}^{\rm min} =&~\underset{\mathbf{W}_i,\,i\in A^{(l)}}{\rm min} \sum_{i\in A^{(l)}} {\rm tr}(\mathbf{W}_i)\\
{\rm s.t.}:&~ {\rm tr}(\mathbf{R}_{i} \mathbf{W}_i)-\gamma_i \sum_{\substack{k\in A^{(l)}\\ k\neq i}} {\rm tr}(\mathbf{R}_{i} \mathbf{W}_{k})\geq \gamma_i \sigma^2,\,i\in A^{(l)},\\
& ~\mathbf{W}_{i} \succeq 0, ~i\in A^{(l)}.
\end{eqn}
Interestingly, it turns out that the SDR form (\ref{eq:socpUEs}) and the original problem (\ref{eq:qcqp2}) are equivalent \cite{Gershman2010ConvxOptBeamform}. Therefore, the solution to the SDR problem outputs rank-1 matrices. However, it should be noted that, in general, an SDR problem gives a lower-bound on the optimal objective function.
\subsubsection{UDD-Based Solution}
While the SDR method gives out the optimal solution for the problem in (\ref{eq:socpUEs}), a more efficient solver is a fast iterative algorithm based on the uplink-downlink duality \cite{Schubert2004BeamformSINRconstraints}. The optimization in the SDR method is performed over the $M$-by-$M$ $\mathbf{W}_i$ matrices, which have more unknowns than the original $M$-element beamforming vectors, i.e., the $\mathbf{w}_i$'s. Hence, the SDR solution comes at the cost of a relatively high computational complexity. To take advantage of the UDD method, we re-write the problem (\ref{eq:qcqp2}) by using the normalized beamforming vectors, i.e., $\mathbf{u}_i = {\mathbf{w}}_{i} / \norm{\mathbf{w}_i}$, and putting them in the matrix $\mathbf{U} = (\mathbf{u}_1, \dots, \mathbf{u}_K)$, while denoting $\rho_i=\norm{\mathbf{w}_i}^2$. Accordingly, the problem becomes
\begin{eqn}\label{eq:qcqp3}
P_{A^{(l)}}^{\rm min}=& \underset{\mathbf{U},\,\bm{\rho}}{\rm min} \Vert \bm{\rho} \Vert_{1}\\
{\rm s.t.}:&~\Gamma_i(\mathbf{U},\bm{\rho}) \geq \gamma_i, \Vert \mathbf{u}_i \Vert_{1} = 1,~  i\in A^{(l)},
\end{eqn}%
where $\bm{\rho}=(\rho_1,\dots,\rho_K)^{T}$, and $\Gamma_i(\mathbf{U}, \bm{\rho})$ is obtained by replacing $\mathbf{w}_i=\sqrt{\rho_i} \mathbf{u}_i$ in (\ref{eq:Gammak}). Then, the minimum powers, $\rho_i^{\star}$'s, and the normalized beamforming vectors, $\mathbf{u}_i^{\star}$'s, can be found by applying the algorithm in \cite[Table II]{Schubert2004BeamformSINRconstraints}.
\subsubsection*{{\bf Remark 1}}
{\it In order for problem (\ref{eq:qcqp2}) to always have a solution, the rank of the channel matrix $\mathbf{H}=[\mathbf{h}_1, \dots,\mathbf{h}_{K}]$ should be greater than or equal to the number of UEs. For well-conditioned channels, the latter condition becomes $M\geq K$. Of course, there may be solutions for cases in which  $M < K$, depending on the devices' required SINRs, i.e., the $\gamma_i$'s, and their channels, i.e., the $\mathbf{h}_i$'s.}

A brief interpretation of Remark 1 is as follows. A solution $P_{A^{(l)}}^{\rm min}<\infty$ for (\ref{eq:qcqp3}) depends on the SINR demands as well as the channel coefficients of the users. For example, when two users ($\mathrm{UE}_1$ and $\mathrm{UE}_2$) are exactly beside each other, they will have same channel coefficients, i.e., $\mathbf{h}_1 = \mathbf{h}_2$. In this case, the problem in (\ref{eq:qcqp3}) will have a solution {\it iff} $\gamma_1<1$ and $\gamma_2<1$. Intuitively, in this example, the power signal to a user produces the same amount of interference to the other. So it is required that the users be spatially separated, i.e., each has its own independent beam, in order for the problem to always have a solution. The maximum number of independent beams is equal to the rank of $\mathbf{H}$ \cite{Tse2004WirelessCommEbook}, which for a well-conditioned matrix is equal to $\min\{M, K\}$. Similar to an $M\times M$ point-to-point link which is capable of multiplexing $M$ independent streams, for $M < K$ the maximum number of independent beams is $K$ and, in this case, there might be users with correlated channel coefficients.

With either method, i.e., SDR or UDD, the AP should solve the subproblems in (\ref{eq:qcqp2}) for at most $2^{K}-1$ times to obtain the optimal allocation vector $\mathbf{a}^{\star}$. While using the iterative algorithm in \cite{Schubert2004BeamformSINRconstraints} is faster than solving with the SDR-based method, the exponential dependence of either of the two solution methods on the number of UEs makes them time-consuming, especially when the number of UEs is large. Sub-optimal solvers can be applied to overcome this issue. In subsection \ref{ssec:goodness}, we propose an iterative suboptimal solution with polynomial time complexity. Later in Section \ref{sec:NN}, we propose a DNN-based suboptimal solution which outputs the allocation vector in real time, and is applicable for complex systems where conventional iterative solvers lose their validity.
\subsubsection*{\bf{Remark 2} [{\it Monotonicity}]}
{\it Fixing the beamforming matrix to $\tilde{\mathbf{U}}$, it is clear that by increasing the AP's maximum power $P$, the SINR constraints of all devices are better guaranteed. This is an immediate consequence of $\Gamma_i(\tilde{\mathbf{U}}, \alpha P)\geq \Gamma_i(\tilde{\mathbf{U}}, P)$, where $\alpha > 1$ \cite{Schubert2004BeamformSINRconstraints}.}
\subsection{Wireless Power Transfer Network}\label{ssec:WPT}
In this operating scenario, there are only ER devices, i.e., $\cal K = \cal J$. In this case, the optimization problem (\ref{eq:qcqp1}) boils down to the following separable QCQP problem:
\begin{eqn}\label{eq:opER1}
P_{A^{(l)}}^{\rm min}=& \underset{\mathbf{v}_j,\,j\in A^{(l)}}{\rm min} \sum_{j\in A^{(l)}} \Vert \mathbf{v}_j \Vert^2\\
{\rm s.t.}:~ & Q_j \geq q_j,   j\in A^{(l)},
\end{eqn}
which can be converted to the following SDR program after relaxing the rank-1 constraint:
\begin{eqn}\label{eq:sd_ER}
P_{A^{(l)}}^{\rm min} =&~ \underset{\mathbf{V}_j,\,j\in A^{(l)}}{\rm min} \sum_{j\in A^{(l)}} {\rm tr}(\mathbf{V}_j)\\
{\rm s.t.}:&~\mathrm{tr}(\mathbf{C}_{j} \mathbf{V}_j) \geq q_j,   j\in A^{(l)},\\
&~\mathbf{V}_{j} \succeq 0, ~j\in A^{(l)},
\end{eqn}%
which can be solved using CVX. Noting that in the separable QCQP problem (\ref{eq:sd_ER}), the number of constraints is equal to the number of summation terms of the objective function, the SDR problem is tight, i.e., solving the SDR in (\ref{eq:sd_ER}) is equivalent to solving the original QCQP in (\ref{eq:opER1}). Thus, the solution of the SDR problem results in rank-1 matrices corresponding to the unique optimal beamforming vectors \cite{Luo2010SemidefiniteRelaxation}. Also, it is interesting to note that irrespective of the number of energy constraints, the optimal $\mathbf{V}_j^{\star}$'s are equal to each other, and recalling that the optimal rank-1 matrix $\mathbf{V}_j^{\star} = \mathbf{v}_j^{\star} {\mathbf{v}_j^{\star}}^{H}$, the optimal beamforming vector would be $\mathbf{v}_1^{\star}$ for all ERs in $A^{(l)}$.
\subsubsection*{\bf{Remark 3} [{\it Robustness of the Solution}]}
{\it Problem (\ref{eq:sd_ER}) always has a solution for any $M$, $\mathbf{g}_j$'s, $q_j$'s, and for any number of ERs. This is true because the interference from other devices are beneficial leakage resources of energy, in contrast to the IR-only scenario where the interference from other devices has a destructive effect on the SINR of a specific device.}
\subsection{SWIPT Network}\label{ssec:Back2IER}
\subsubsection*{\bf{Remark 4} [{\it Number of Required Beams}]}
{\it Recall that for the problem in (\ref{eq:socp1}) to have a solution, the number of AP antennas should be at least equal to the number of IRs (cf. Remark 1). The problem would have a solution even if we take $\mathbf{v}_j=\mathbf{0},   j\in\cal{J}$. This is due to the monotonicity of the $\Gamma_i$'s, $  i \in \cal I$ (cf. Remark 2) and the beneficial type of interference for the ERs (cf. Remark 3). Hence, if the energy leakage from the information beams to the ERs satisfy their own demands, then $\mathbf{v}_j$, $j\in\cal J$, will be zero, i.e., no dedicated energy beams are needed to fulfill the ERs' demands. Otherwise, only one energy beam will be needed for all ERs, similar to the case of the ER-only network (cf. subsection \ref{ssec:WPT}). Thus, our network always needs to find a maximum of $I + 1$ beamforming vectors instead of $I+J$.}
\section{Revenue-Maximizing Mechanism}\label{sec:rev_max_mech}
In the previous section, we obtained the feasible allocation set $\mathcal{A}_{\rm F}$. Now, we aim to find the optimal allocation rule $\mathbf{a}^{\star}$ and the optimal payment rule $\mathbf{p}^\star$, which together constitute the Rmax mechanism. To find the Rmax mechanism, it is initially required that the social-welfare maximization (SWmax) mechanism be obtained because the SWmax solution is applied in obtaining the Rmax mechanism \cite{Hartline2013MechanismDesign_Ebook}. Next, we obtain the SWmax mechanism, based on which and by applying Myerson's lemma, we find the Rmax mechanism in subsection \ref{ssec:RMaxMyerson}.
\subsection{Social-Welfare Maximization---VCG Auction}\label{ssec:SWmax}
Dominant strategy incentive compatible (DSIC) auctions are those desired auctions in which the seller needs no strategy, i.e., need not know the valuation distributions of the UEs to design the auction; and each UE, independent of other agents' bidding strategies, should play truthfully to maximize its own benefit, i.e., $b_k = \nu_k$ for $k\in\cal K$. For single-parameter environments, the SWmax mechanism, which will be obtained shortly, is DSIC and is often called Vickrey-Clarke-Groves (VCG) mechanism in the auction literature~\cite{Nisan2007AlgorithmicGameTheory_Ebook}.
\subsubsection{Optimal Allocation Rule for the SWmax Mechanism}
The optimal allocation vector $\mathbf{a}^{\star}=\mathrm{argmax}_{\mathbf{a}\in\mathcal{A}_{\rm F}} S(\mathbf{b}, \mathbf{a})$ corresponds to the feasible set of users that yields the largest sum of bid values. In other words, given the bid profile $\mathbf{b}$, the maximum social welfare $\mathfrak{S}(\mathbf{b})$ which is equal to $S(\mathbf{b, a^\star})$ is simply found by looking up the table of all social-welfare values corresponding to all allocation vectors in the feasible set $\mathcal{A}_F$ and selecting the maximum.
\subsubsection{Optimal Payment Rule for the SWmax Mechanism}
According to \cite[Lemma~3.1]{Hartline2013MechanismDesign_Ebook}, for each $\mathrm{UE}_k, k\in\cal K$, all bid values of other UEs, i.e., $\mathbf{b}_{-k}$, is a non-decreasing step function in terms of $b_k$. The critical value of this step function is $\tilde{b}_k = \mathfrak{S}(0, \mathbf{b}_{-k}) - \mathfrak{S}_{-k}(\infty, \mathbf{b}_{-k})$, where $\mathfrak{S}(0,\mathbf{b}_{-k})$ is the optimal social welfare from UEs other than $\mathrm{UE}_k$ assuming the latter is \textit{not} served; and $\mathfrak{S}_{-k}(\infty, \mathbf{b}_{-k})$ is the optimal social welfare from UEs other than $\mathrm{UE}_k$ assuming that $\mathrm{UE}_k$ is served \cite{Hartline2013MechanismDesign_Ebook}. Consequently, the AP's SWmax mechanism, based on the bid profile $\mathbf{b}$ from the UEs, is described by \cite{Hartline2013MechanismDesign_Ebook}
\begin{subequations}\label{eqs:SMap}
\begin{equation}\label{eq:SMa}
\mathbf{a}^{\star}=\underset{\mathbf{a}\in\mathcal{A}_\mathrm{F}}{\rm argmax}~S(\mathbf{b, a}),
\end{equation}
\begin{equation}\label{eq:SMp}
p_k^{\star} = \begin{cases} \tilde{b}_k & \text{if } a_k^{\star} = 1\\0 & \text{if } a_k^{\star} = 0\end{cases},\quad k\in\cal K,
\end{equation}
\end{subequations}
where (\ref{eq:SMp}) is the optimal payment rule.

Here, it is worth noting that social welfare maximization is singular among objectives in that there is a single mechanism that is optimal regardless of the distributional assumptions for the agents' valuations. In fact, the agents' incentives are already aligned with the seller's objective, and one only needs to derive the appropriate payments, i.e., the critical values. For general objectives, e.g., revenue maximization which will be obtained in the next subsection, the optimal mechanism is distribution-dependent!
\subsection{Revenue Maximization Mechanism---Mayerson Mechanism}\label{ssec:RMaxMyerson}
We place two standard assumptions on our mechanisms: (i) they are individually rational, meaning that no agent has negative expected utility for taking part in the auction; and (ii) agents who do not win, pay nothing, i.e., $a_k=0 \rightarrow p_k=0$. A mechanism is DSIC (also called truthful) \textit{in expectation} {\it iff} $\mathbb{E}[u_{k}(\nu_k, \mathbf{b}_{-k})]\geq \mathbb{E}[u_k(b_i, \mathbf{b}_{-k})]$, which for single parameter environments translates into a mechanism having the following conditions \cite{Nisan2007AlgorithmicGameTheory_Ebook}: (i) $a_k(b_k, \mathbf{b}_{-k})$ is \textit{monotone non-decreasing} in $b_k$ assuming $\mathbf{b}_{-k}$ is fixed,\footnote{With a non-decreasing monotone allocation rule, bidding less does not cause a bidder to get more of the commodity.} and (ii) $p_k(b_k,\mathbf{b}_{-k})= b_k a_k(b_k,\mathbf{b}_{-k}) - \int_{0}^{b_k} a_k(z,\mathbf{b}_{-k}) {\rm d}z$ (a.k.a. Myerson's payment identity). Thus, once the allocation rule is fixed, the payment rule is found by applying the Myerson's payment identity.

As aforementioned, private valuation $\nu_k, k\in \mathcal{K}$, is drawn from the distribution $\mathcal{F}_k(\nu_k)$ with density function $f_{k}(\nu_k)$. By taking expectation of both sides of the the Myerson's payment identity, and summing over all agents, we end up with the following key relation
\begin{equation}\label{eq:Evpiv}
\mathbb{E}_{\bm{\nu}}\left [ \sum_{k\in\mathcal{K}} p_k(\bm{\nu}) \right ] = \mathbb{E}_{\bm{\nu}}\left [\sum_{k\in\mathcal{K}} \phi_k(\nu_k)a_k(\bm{\nu}) \right ],
\end{equation}
where $\phi_k(\nu_k)=\nu_k - \frac{1-\mathcal{F}_k(\nu_k)}{f_k(\nu_k)}$ is the \textit{virtual valuation} for every $\mathrm{UE}_k, k\in\cal K$.

Referring to $\sum_{k\in\mathcal{K}} \phi_k(\nu_k)a_k(\bm{\nu})$ as the \textit{virtual social welfare} of an auction on the valuation profile $\bm{\nu}$, (\ref{eq:Evpiv}) states that the expected revenue equals the expected virtual social welfare. Thus, the virtual social-welfare-maximizing (VSM) allocation rule is one which chooses the feasible allocation that maximizes the virtual social welfare $\sum_{k\in\mathcal{K}} \phi_k(\nu_k)a_k(\bm{\nu})$ for each valuation profile $\bm{\nu}$. For the VSM mechanism to be truthful, the obtained allocation rule has to be monotone non-decreasing, which holds true when the virtual valuations $\phi_k(v_k)$ are monotone non-decreasing.\footnote{A sufficient condition for monotone virtual valuations is implied by the monotone hazard rate assumption. If the hazard rate of distribution $\mathcal{F}_k$, defined as $f_k(\nu_k)/(1-\mathcal{F}_{k}(\nu_k))$, is monotone non-decreasing, then the virtual valuations are
monotone nondecreasing as well.}

Therefore, the optimal Rmax mechanism, a.k.a. Mayerson mechanism, can be described as follows \cite{Hartline2013MechanismDesign_Ebook}
\begin{subequations}\label{eq:VSMap}
\begin{equation}\label{eq:VSMa}
(\mathbf{a}^{\star}, \mathbf{p}^{\prime})= \mathrm{VCG}^{\prime}(\mathbf{b}^{\prime})
\end{equation}
\begin{equation}\label{eq:VSMp}
p_k^{\star} = \begin{cases} \phi_k^{-1}(p_k^{\prime}) & \text{if } a_k^{\star}=1\\0 & \text{if } a_k^{\star}=0\end{cases},~~k\in\cal K,
\end{equation}
\end{subequations}%
where $b_k^{\prime}=\phi_k(b_k), k\in \mathcal{K},$ are the elements of the virtual bid profile $\mathbf{b}^{\prime}$, and where $\mathrm{VCG}^{\prime}(\mathbf{b}^{\prime})$ is the generalized version of the SWmax mechanism function described in (\ref{eqs:SMap}). In fact, since the virtual bids can have negative values, the generalized SWmax mechanism, which takes as input the virtual bid profile $\mathbf{b}^{\prime}$ and outputs $(\mathbf{a}^{\star}, \mathbf{p}^{\prime})$, is defined as follows:
\begin{subequations}\label{eqs:SMap2}
\begin{equation}\label{eq:SMa2}
\mathbf{a}^{\star}=\begin{cases}\underset{\mathbf{a}\in\mathcal{A}_\mathrm{F}}{\rm argmax}~ S(\mathbf{b^{\prime}, a}), & \text{if } S(\mathbf{b^{\prime}, a^{\star}}) > 0,\\
\mathbf{0}, & \text{Otherwise},\end{cases}
\end{equation}
\begin{equation}\label{eq:SMp2}
p_k^{\prime} = \begin{cases} \mathfrak{S}(-\infty, \mathbf{b}_{-k}^{\prime}) - S_{-k}^{\star}(\infty, \mathbf{b}_{-k}^{\prime}), & \text{if } a_k^{\star} = 1\\0, & \text{if } a_k^{\star} = 0\end{cases},~k\in\cal K,
\end{equation}
\end{subequations}
where $\mathfrak{S}(-\infty, \mathbf{b}_{-k}^{\prime})$ is the optimal social welfare from UEs other than $\mathrm{UE}_k$, assuming the latter is \textit{not} served; and $S_{-k}^{\star}(\infty, \mathbf{b}_{-k}^{\prime})$ is the optimal social welfare from UEs other than $\mathrm{UE}_{k}$, assuming that $\mathrm{UE}_k$ is served. Hence, $p_k^{\prime} \geq 0$ for $k\in \cal K$.

The pair $(\mathbf{a}^{\star}, \mathbf{p}^{\star})$ that is obtained from (\ref{eq:VSMa}) and (\ref{eq:VSMp}), constitutes the Rmax mechanism which can be described in algorithmic steps as follows: (i) given the bid profile $\mathbf{b}$ and the distributions $\mathcal{F}_{k}, k\in\cal K$, compute virtual bids $b_k^{\prime}=\phi_{k}(b_k), k\in\cal K$, (ii) run $\mathrm{VCG}^{\prime}$ described in (\ref{eqs:SMap2}) on the virtual bids $\mathbf{b}^{\prime}$ to get $\mathbf{a}^{\star}$ and $\mathbf{p}^{\prime}$, (iii) compute $p_k^{\star}$ by applying (\ref{eq:VSMp}), and (iv) output ($\mathbf{a^{\star}, p^{\star}}$).

\begin{figure}
\centering
\includegraphics[width=0.7\linewidth]{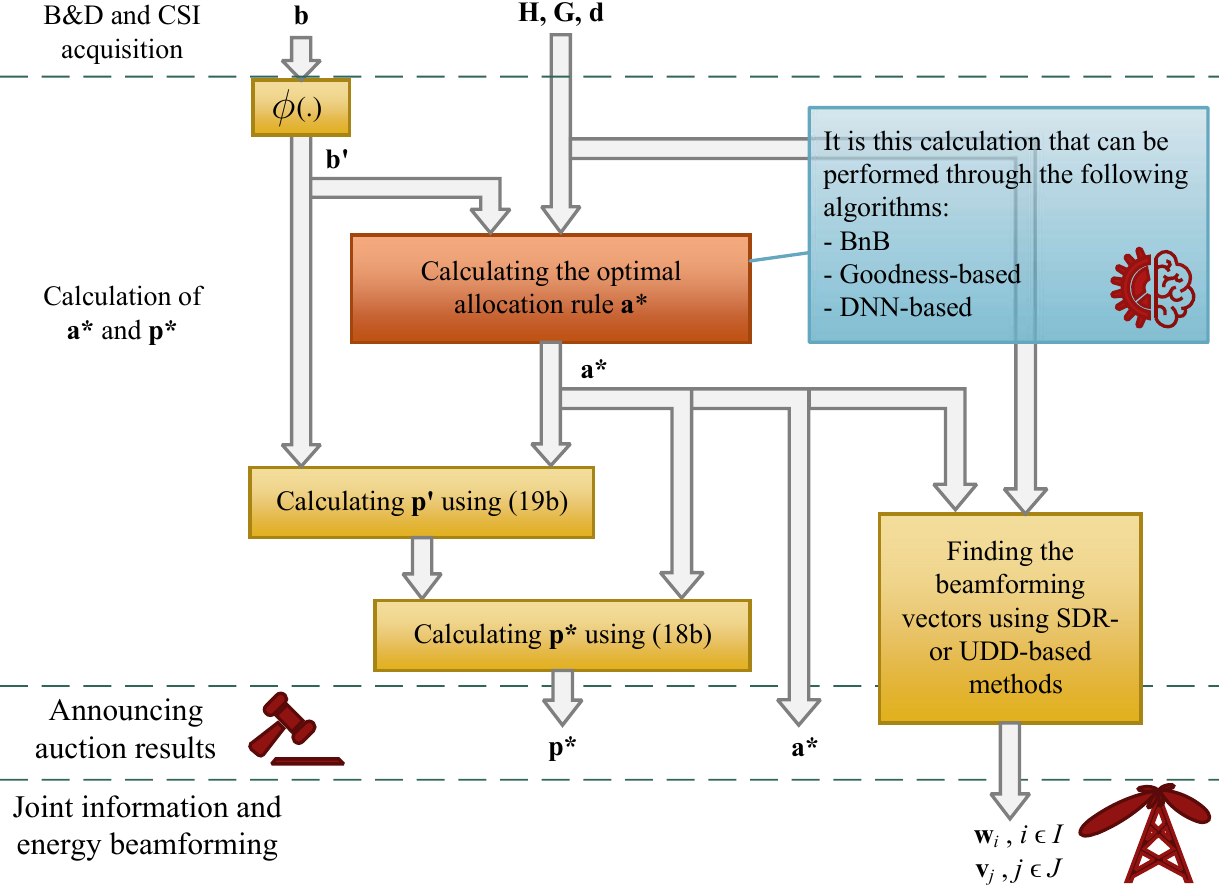}
\caption{Flow diagram of finding the optimal revenue maximization mechanism.}\label{fig:FlowDiagram}
\end{figure}

The flow diagram of Fig. \ref{fig:FlowDiagram} illustrates the whole process of finding the optimal mechanism.
\section{Iterative Algorithms for Finding the Allocation Rule}\label{sec:Algos}
\subsection{Branch-and-Bound Algorithm}
Let $\mathcal{L}_k(\mathcal{K}),\, k\in\{0,1,\dots,|\mathcal{K}|\}$, hold all the $k$-element subsets of set $\cal K$, where $|\mathcal{K}|$ is the cardinality of $\mathcal{K}$. Mathematically speaking, we have $\mathcal{L}_k(\mathcal{K})=\{s|s\subset \mathcal{P}(\mathcal{K}), \, |s| = k\}$ where $\mathcal{P}(\mathcal{K})$ is the power set of $\mathcal{K}$, and $|\mathcal{L}_k(\cal K)|=(\substack{|\mathcal{K}|\\k})$ is the cardinality of $\mathcal{L}_k(\cal K)$.
Considering these notations, the efficient BnB algorithm, i.e., Algorithm \ref{alg:find_aDaE}, can be applied to find the optimal allocation rule $\mathbf{a}^{\star}=(\substack{\mathbf{a}_{\rm IR}^{\star}\\\mathbf{a}_{\rm ER}^{\star}})$. One should note that the first breadth level of the search tree of the proposed BnB algorithm starts with the largest subset of users, i.e., $\left (\mathcal{L}_{|\mathcal{I}|}(\mathcal{I}), \mathcal{L}_{|\mathcal{J}|}(\mathcal{J})\right )$. In other words, the branches are formed based on the \textit{exclusion} of a user rather than \textit{inclusion}. This approach is inspired by the fact that all feasible subsets are downward-closed.

\begin{algorithm}[h]
\caption{The efficient Branch-and-Bound algorithm for finding $\mathbf{a^\star}$.}\label{alg:find_aDaE}
\begin{algorithmic}[1]
\State \textit{Initialize}: $A_{\rm IR}=\{\}$, $A_{\rm ER}=\{\}$, $U=0$, $\mathcal{O}=\left \{\left (\{\}, \{\}\right )\right \}$
\State Compute $b_k^{\prime}, k \in \cal K$, using the corresponding valuation functions
\For{$m=I$ to $0$}
	\For{each subset $X$ in $\mathcal{L}_m(\mathcal{I})$}
		\For{$n=J$ to $0$}
			\For{each subset $Y$ in $\mathcal{L}_n(\mathcal{J})$}
				\If{$\{(X, Y)\}\nsubseteq \mathcal{O}$}
					\State Find $P^{\prime}$ by solving (\ref{eq:socp1}) for the subset $\{(X, Y)\}$ via CVX
					 \State $U^{\prime} \leftarrow \sum_{i\in X} 									 	b_{i}^{\prime}+ \sum_{j\in Y} b_{j}^{\prime}$
					 \If{$U^{\prime} > U$ \textbf{and} $P^{\prime} < P$}
					 	\State $A_{\rm IR} \leftarrow X$
					 	\State $A_{\rm ER} \leftarrow Y$
					 	\State $U \leftarrow U^{\prime}$
					 	\State \Call{UpdateO}{$X, Y$}
					 \EndIf
				\EndIf
			\EndFor
		\EndFor
	\EndFor
\EndFor
\State Output $(\substack{\mathbf{a}_{\rm IR}\\\mathbf{a}_{\rm ER}})$ corresponding to $\{A_{\rm IR}, A_{\rm ER}\}$ as the optimal allocation rule $\mathbf{a}^{\star}$
\Procedure{UpdateO}{$X, Y$}
	\For{$i=|X|$ to $0$}
		\For{each subset $X^{\prime}$ in $\mathcal{L}_i(X)$}
			\For{$j=|Y|$ to $0$}
				\For{each subset $Y^{\prime}$ in $\mathcal{L}_j(Y)$}
					 \If{$\{(X^{\prime}, Y^{\prime})\}\nsubseteq \mathcal{O}$}
						\State Add $(X^{\prime}, Y^{\prime})$ to $\mathcal{O}$
					 \EndIf
				\EndFor
			\EndFor
		\EndFor
	\EndFor
\EndProcedure
\end{algorithmic}
\end{algorithm}
\subsection{Heuristic Sub-Optimal Iterative Solutions}\label{ssec:goodness}
The BnB algorithm to find the feasible set $\mathcal{A}_{\rm F}$ has exponential time complexity in terms of the number of devices, i.e., $K$. Apart from the very simplistic method of random allocation, next we present an iterative heuristic suboptimal solution which has polynomial-time complexity. Thereafter, we will propose a real-time DNN-based sub-optimal solution.

This allocation strategy is based on the the \textit{goodness} factors $\lambda_i=\sigma_{\mathbf{h}_i}^{2} b_i^{\prime}/\gamma_i$ for the IRs, $i\in\cal I$, and $\mu_j=\sigma_{\mathbf{h}_j}^{2}  b_j^{\prime}/q_j$ for the ERs, $j\in\cal J$.
In this method, the AP starts with the largest subset $\mathcal{L}_{K}(\cal K)$ and obtains the solution of the optimization problem (\ref{eq:qcqp1}). If the subset is found unfeasible, the AP would drop the device (either IR or ER) having the least goodness value. Then, the AP checks the feasibility of the remaining subset. This procedure is repeated until the first feasible set is obtained. This method converges to a suboptimal solution in a maximum of $K -1$ running times of the iterative optimization method, either SDR-based or UDD-based. The goodness-based algorithm is depicted in Algorithm \ref{alg:goodness}.

The goodness factor is more meaningful when there are only IRs or ERs in the network. For the WIT scenario investigated in subsection \ref{ssec:WIT}, the goodness-based (here called $\lambda$-based) algorithm converges in $I - 1$ runs of the problem solver algorithm, e.g., the efficient UDD-based method, in the worst case. In the WPT scenario discussed in subsection \ref{ssec:WPT}, the goodness-based (called $\mu$-based here) algorithm converges in $J - 1$ runs of the applied problem solver, e.g., SDR-based standard tool CVX, at worst case.
\begin{algorithm}[h]
\caption{The heuristic goodness-based algorithm for finding $\mathbf{a^\star}$.}\label{alg:goodness}
\begin{algorithmic}[1]
\State \textit{Initialize}: $A=\{\}$; $U=0$; $\mathbf{a}^{\star} = \mathbf{0}$
\State Compute $b_k^{\prime}, k \in \cal K$, using the corresponding valuation functions
\State Make the ordered set $\mathcal{K}^{\prime}$ sorted increasingly based on $\lambda_i=\sigma_{\mathbf{h}_i}^{2} b_i^{\prime}/\gamma_i$ and $\mu_j=\sigma_{\mathbf{h}_j}^{2} b_j^{\prime}/q_j$
\While{$\mathcal{K}^{\prime}\neq \emptyset$}
\State Find the minimum power $P^{\prime}$ by solving (\ref{eq:socp1}) for the set $\mathcal{K}^{\prime}$ via CVX
\State $U^{\prime} \leftarrow \sum_{k\in \mathcal{K}^{\prime}} b_k^{\prime}$
\If{$U^{\prime} > U$ \textbf{and} $P^{\prime} < P$}
\State Break
\Else
\State Drop the first element of $\mathcal{K}^{\prime}$ with the least value of goodness factor.
\EndIf
\EndWhile
\State Output $\mathbf{a}^{\star} = (\substack{\mathbf{a}_{\rm IR}\\\mathbf{a}_{\rm ER}})$ corresponding to $\mathcal{K}^{\prime}$ as the optimal allocation rule
\end{algorithmic}
\end{algorithm}
\section{Deep Learning Based Allocation Rule}\label{sec:NN}
As observed in Section \ref{sec:FeasibleSet}, the time-greedy part of designing the optimal mechanism for the SWIPT network under consideration is due to the many runs of the optimization methods needed to find the allocation rule. In fact, once the optimal allocation rule is found, finding the optimal payment rule is straightforward by applying (\ref{eq:VSMap}) and (\ref{eqs:SMap2}). One solution to overcome the complexity of finding the allocation vector is through the use of DNNs.

To solve $\mathbf{a}^{\star}=\mathrm{argmax}_{\mathbf{a}\in\mathcal{A}_{\rm F}} S(\mathbf{b}^{\prime}, \mathbf{a})$ via deep learning, we observe that it can be regarded as an unknown function mapping from the ensemble of the network parameters of interest, i.e., $\mathbf{b}^{\prime}$, $\mathbf{d}$, and $\mathbf{M}$, where $\mathbf{M}=[\mathbf{H}, \mathbf{G}]$ with $\mathbf{H}=[\mathbf{h}_{1}, \dots, \mathbf{h}_{I}]$ and $\mathbf{G}=[\mathbf{g}_1, \dots, \mathbf{g}_J]$, to the corresponding optimal allocation rule $\mathbf {a}^{\star}=(\substack{\mathbf{a}_{\rm IR}^{\star}\\\mathbf{a}_{\rm  ER}^{\star}})$. Note that the mapping depends on $\mathbf{h}_k$'s and $d_k$'s through $\mathcal{A}_{\rm F}$. Indeed, DNNs can be viewed as \textit{universal approximators}: if properly trained, they are able to learn the input-output relationship between the parameters and the desired allocation vector. This means that we can optimize a desired performance function for given parameters without having to explicitly solve any optimization problem via SDR, UDD, or any other iterative optimization method, but rather letting the DNN compute the allocation vector.
\subsection{The Proposed Deep Neural Network Architecture}
After trying several core architectures, such as fully-connected neural network (FcNN), convolutional neural network (CNN), and residual neural network (ResNet), the FcNN model showed the best performance in terms of accuracy.
\begin{figure}[h]
\centering
\includegraphics[width=0.313\linewidth]{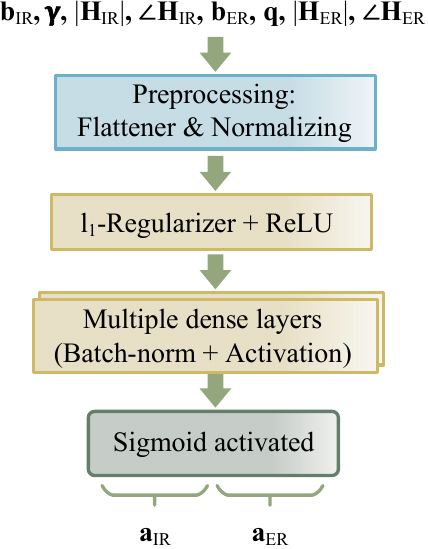}
\caption{The DNN used for  finding the allocation vector.}\label{fig:NN1}
\end{figure}

The schematic of the proposed DNN architecture is depicted in Fig.~\ref{fig:NN1}. The input to the DNN is $[\mathbf{b}_{\rm IR}^{\prime}, \bm{\gamma}, |\mathbf{H}|, \angle\mathbf{H}, \mathbf{b}_{\rm ER}^{\prime}, \mathbf{q}, |\mathbf{G}|, \angle\mathbf{G}]$, where $|\mathbf{H}|$ ($|\mathbf{G}|$) and $\angle{\mathbf{H}}$ ($\angle{\mathbf{G}}$) denote matrices holding the absolute value and angle of the complex elements of matrix $\mathbf{H}$ ($\mathbf{G}$). The input data is fed to a preprocessing unit composed of four main operations: {\it transforming}, {\it sorting}, {\it flattening}, and {\it normalizing}. In transforming, we use the criteria employed in the proposed heuristic method to replace the values of the virtual bids with their corresponding goodness factors. Then, the resulting goodness factors are sorted in ascending (or descending) order. The introduction of these two preprocessing operations makes the training process converge faster and result in a more accurate network at the end of the training. With flattening, the input data are placed in a column vector, and by the normalization operation the flattened data are shifted and scaled according to standard Normal distribution. To explain the latter more, for each input data sample $x$, the normalizing unit operates $\tfrac{x - \mu_x}{\sigma_x}$ on $x$, where $\mu_x$ and $\sigma_x^2$ are the standard mean and variance functions, both of which are obtained from the training data samples. The output layer has $K=I+J$ nodes, which corresponds to the dimension of the allocation vector $\mathbf{a}=(\substack{\mathbf{a}_{\rm IR}\\\mathbf{a}_{\rm ER}})$. The output layer, after being activated by the sigmoid (logistic) function, will take values between $0$ and $1$. In fact, since our network is a multi-label multi-class classifier, the activation function of the output layer should be sigmoid function $f(x)=\tfrac{1}{1+\exp(-x)}$ that maps the summation node value $x\in \mathbb{R}$ to $(0, 1)$.

When using the trained DNN for prediction, we should revert the sorted data back to its original form based on the sorting index set obtained in the sorting operation of the preprocessing unit. Also, we round this output to only take binary values, since the allocation vector is a binary vector.

To train the proposed DNN, we need to populate the training dataset and the corresponding labels. To this end, we solve the corresponding optimization problems for many network realizations using standard semidefinite problem solvers like CVX \cite{Grant2014CVX} in the proposed efficient BnB algorithm. We used the {\it TensorFlow} interface \cite{Abadi2015Tensorflow} for building and training our DNN model. We tuned the hyper-parameters of the DNN using the recently released hyper-parameter optimization framework {\it Keras-tuner} \cite{Kerastuner2019}, which tries a preset number of trials looking for the best possible set of hyper-parameters with built-in search algorithms.
\subsection{Computational Complexity}
%
The iterative-based algorithms, SDR- and UDD-based, have exponential time complexity. In Table \ref{tb:complexity}, the computational complexity of the proposed goodness-based DNN-based methods and the conventional SDR-based and UDD-based methods are compared. In computing complexity, it is assumed that for the SDR-based case the interior-point algorithm \cite{Vandenberghe2005SDPInteriorPoint} is used for each branch $k\in\{1,\dots,2^K-1\}$ of the proposed BnB algorithm, i.e., Algorithm \ref{alg:find_aDaE}, and that for the UDD-based method the iterative algorithm in \cite{Schubert2004BeamformSINRconstraints} is used in each branch of the BnB algorithm for the case of the WIT system. As can be seen from Table \ref{tb:complexity}, in terms of the number of antennas, $M$, the conventional methods as well as the goodness-based method have polynomial time complexity, whereas the proposed DNN-based method has linear time complexity.
Importantly, while the conventional algorithms both have exponential time complexity in terms of the number UEs, $K$, the heuristic method has polynomial time complexity, and the DNN-based method has linear complexity.

\begin{table}[h]
\centering
\caption{Comparison of the Computational Complexity}\label{tb:complexity}
\begin{tabular}{c|c|c|c}%
\textbf{SDR} & \textbf{UDD} & \textbf{Goodness} & \textbf{DNN}\\
\hline\hline
$\mathcal{O}\left(K (M^3+K M^2)~2^K\right)$ & $\mathcal{O}\left (K(M^3 + KM) ~2^K\right )$ & $\mathcal{O}\left (K^2(M^3 + KM)\right )$ & $\mathcal{O}(M K)$
\end{tabular}
\vspace{-0.2in}
\end{table}
\section{Performance Evaluation and Discussion}\label{sec:PerformanceEvaluation}
We evaluate the accuracy of the proposed DNN architecture in predicting the exact allocation vectors. We instantiate our DNN from the architecture presented in Fig. \ref{fig:NN1} targeted for a system operation where transmissions undergo Rayleigh fading. Specifically, $\mathbf{h}_i\sim \mathcal{CN}(\mathbf{0}, \sigma_{\mathbf{h}_i}^2 \mathbf{I}_M)$ where $\sigma_{\mathbf{h}_i}^2\sim\mathcal{U}(-80, -60)\,\rm dB,$ $  i\in \cal I$, and $\mathbf{g}_j\sim \mathcal{CN}(\mathbf{0}, \sigma_{\mathbf{g}_j}^2 \mathbf{I}_M)$ where $\sigma_{\mathbf{g}_j}^2\sim\mathcal{U}(-60, -40)\,\rm dB,$ $  j\in \cal J$. Unless otherwise stated, the SWIPT network parameters are as follows: $P_{\rm max} = 3\,\rm W$, $I=4$, $J=2$, $M=8$, $\sigma_{i}^2 = -50\,\rm dBm$, $i\in \cal I$, $b_k\sim\mathcal{U}(0.1, 1)$ for $k\in\cal K$, $\gamma_i\sim\mathcal{U}(5, 35)\,\rm dB$ for $i\in \cal I$, and $q_j\sim\mathcal{U}(-20, 0)\,{\rm dB}$ for $j\in \cal J$. ${\cal U}(a, b)$ denotes the uniform distribution in the interval $(a,b)$.

\begin{table}[h]
\centering
\caption{Layout of the proposed DNN architecture (\# trainable parameters: 201,454; \# non-trainable parameters: 1,376)}\label{tb:NNparams}
\begin{tabular}{l|c}\hline
\hspace*{2cm}Layer & Output Dimension\\
\hline\hline
Input (after preprocessing) & 108\\
Dense + Regularizer + tanh & 200\\
Dense + Batch normalization + tanh & 296\\
Dense + Batch normalization + ReLU & 392\\
Dense + sigmoid & 6\\\hline
\end{tabular}
\end{table}

\begin{figure}[h]
\centering
\subfloat[]{\includegraphics[width=0.313\linewidth]{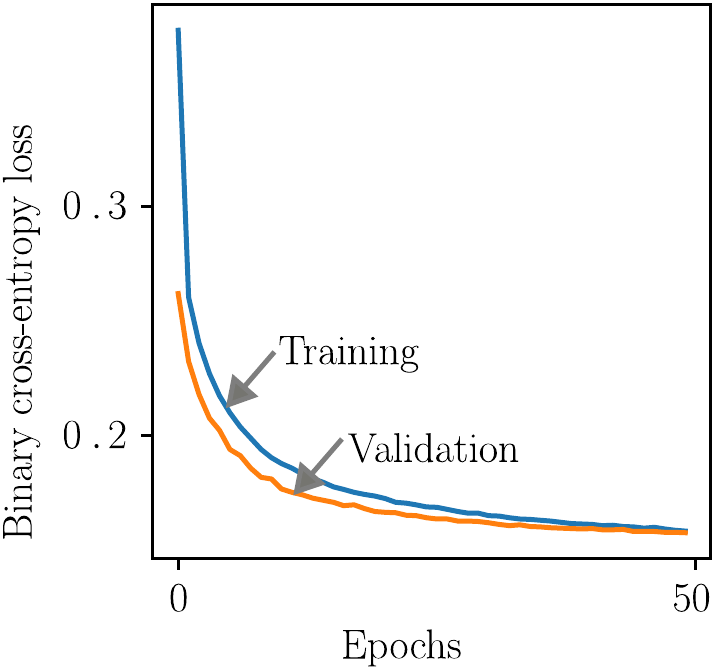}\label{fig:mse}}
\hspace*{0.2in}
\subfloat[]{\includegraphics[width=0.313\linewidth]{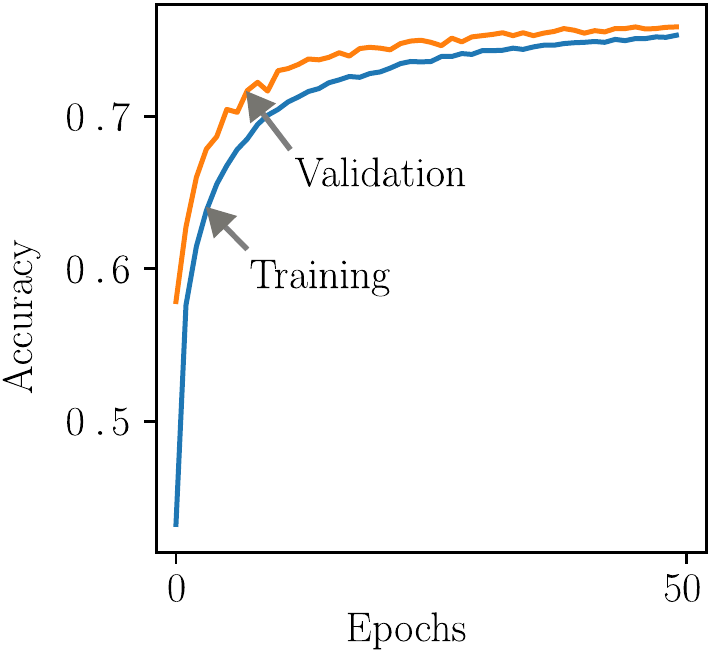}\label{fig:accuracy}}
\caption{(a) The binary cross-entropy metric versus each epoch of training the DNN; (b) The accuracy metric versus each epoch of training the DNN.}\label{fig:bceacc}
\label{fig:accuracy-mse}
\end{figure}

\begin{figure}[h]
\centering
\subfloat[]{\includegraphics[width=0.313\linewidth]{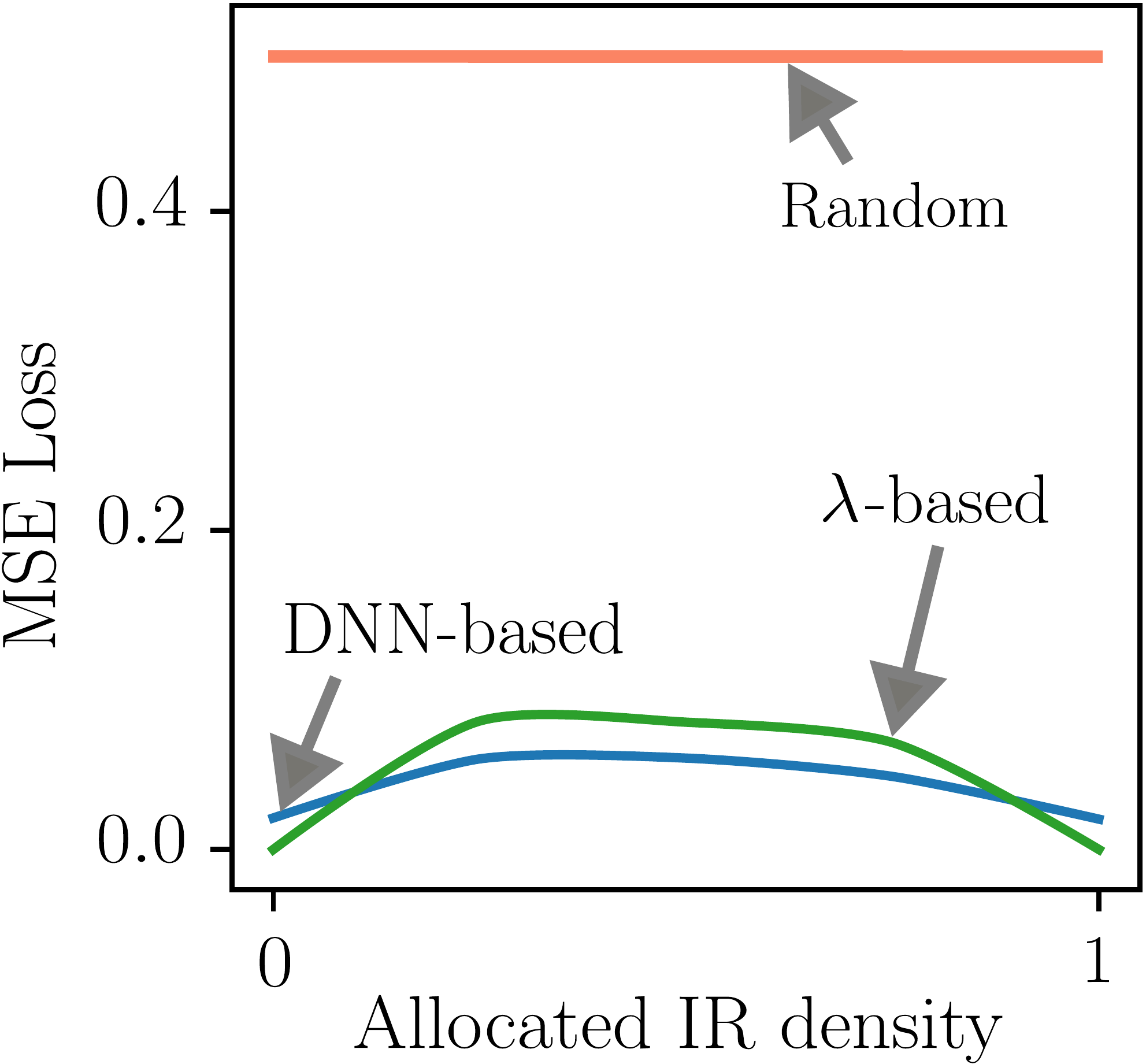}\label{fig:mse_DNNvsLambda}}
\hspace*{0.2in}
\subfloat[]{\includegraphics[width=0.313\linewidth]{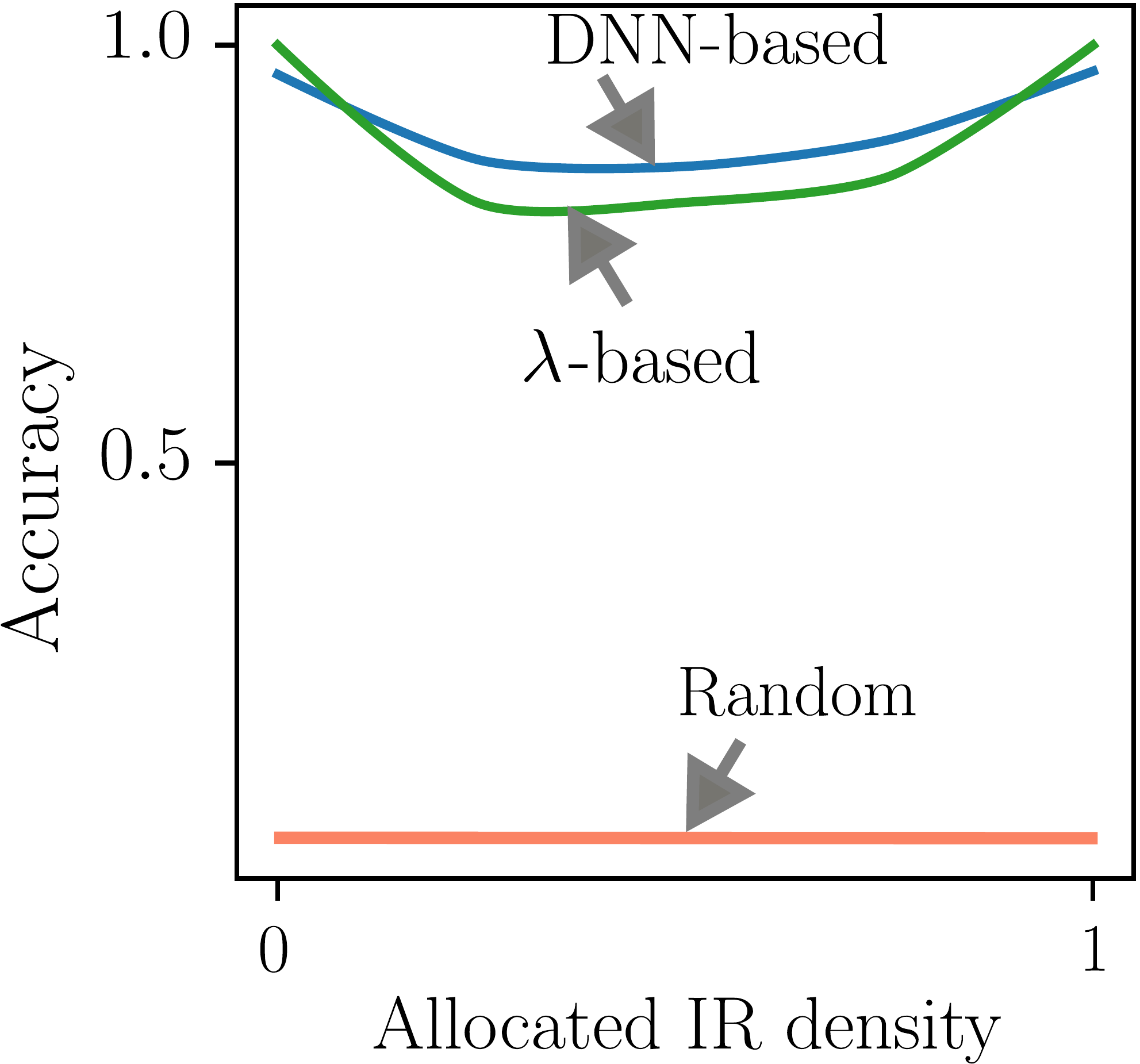}\label{fig:acc_DNNvsLambda}}
\caption{Comparison of the DNN-based and $\lambda$-based methods: (a) The mean square error; (b) The exact accuracy.}
\label{figs:DNNvsLambda}
\end{figure}

\begin{figure}[h]
\centering
\subfloat[]{\includegraphics[width=0.313\linewidth]{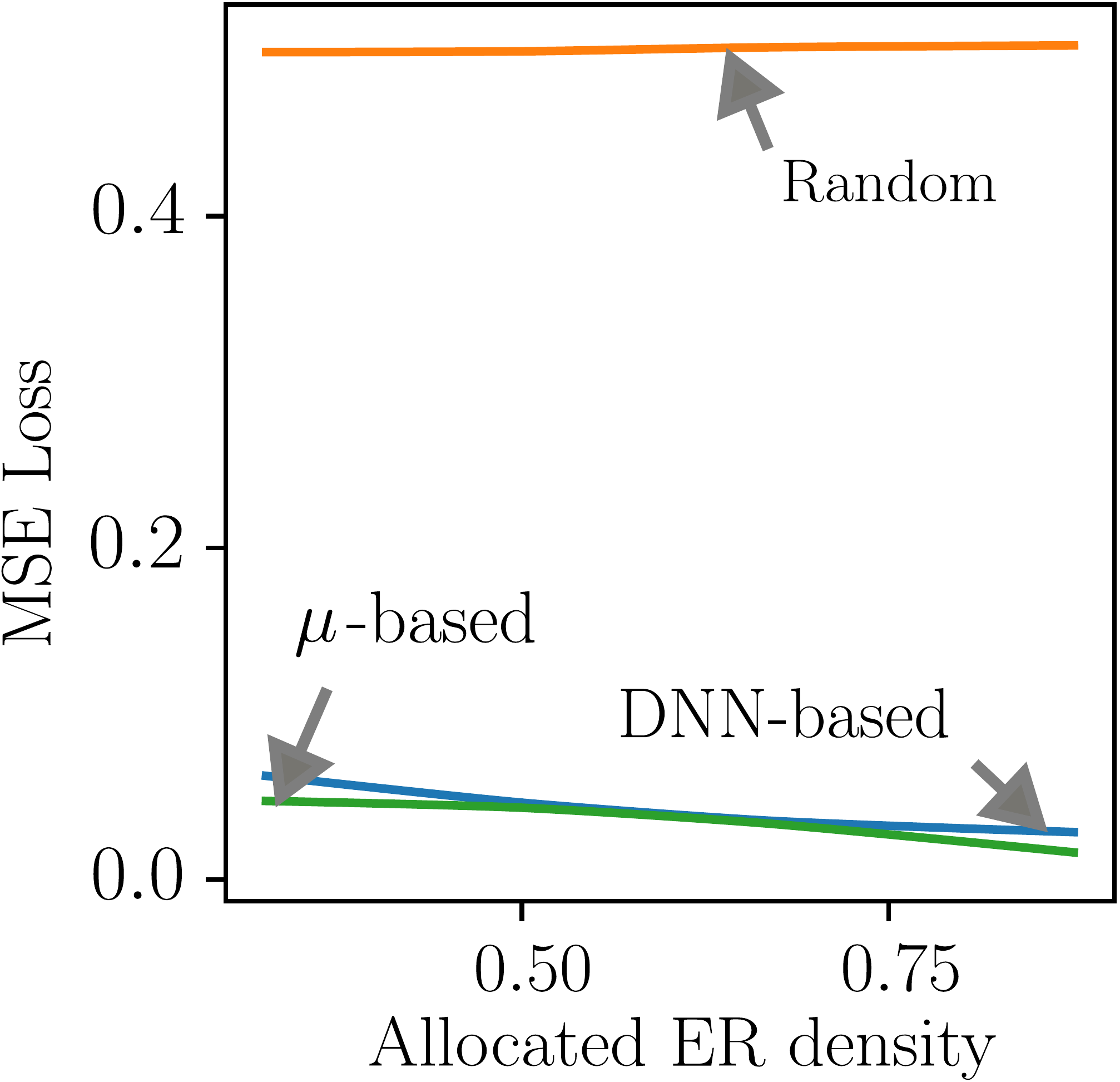}\label{fig:mse_DNNvsMu}}
\hspace*{0.2in}
\subfloat[]{\includegraphics[width=0.313\linewidth]{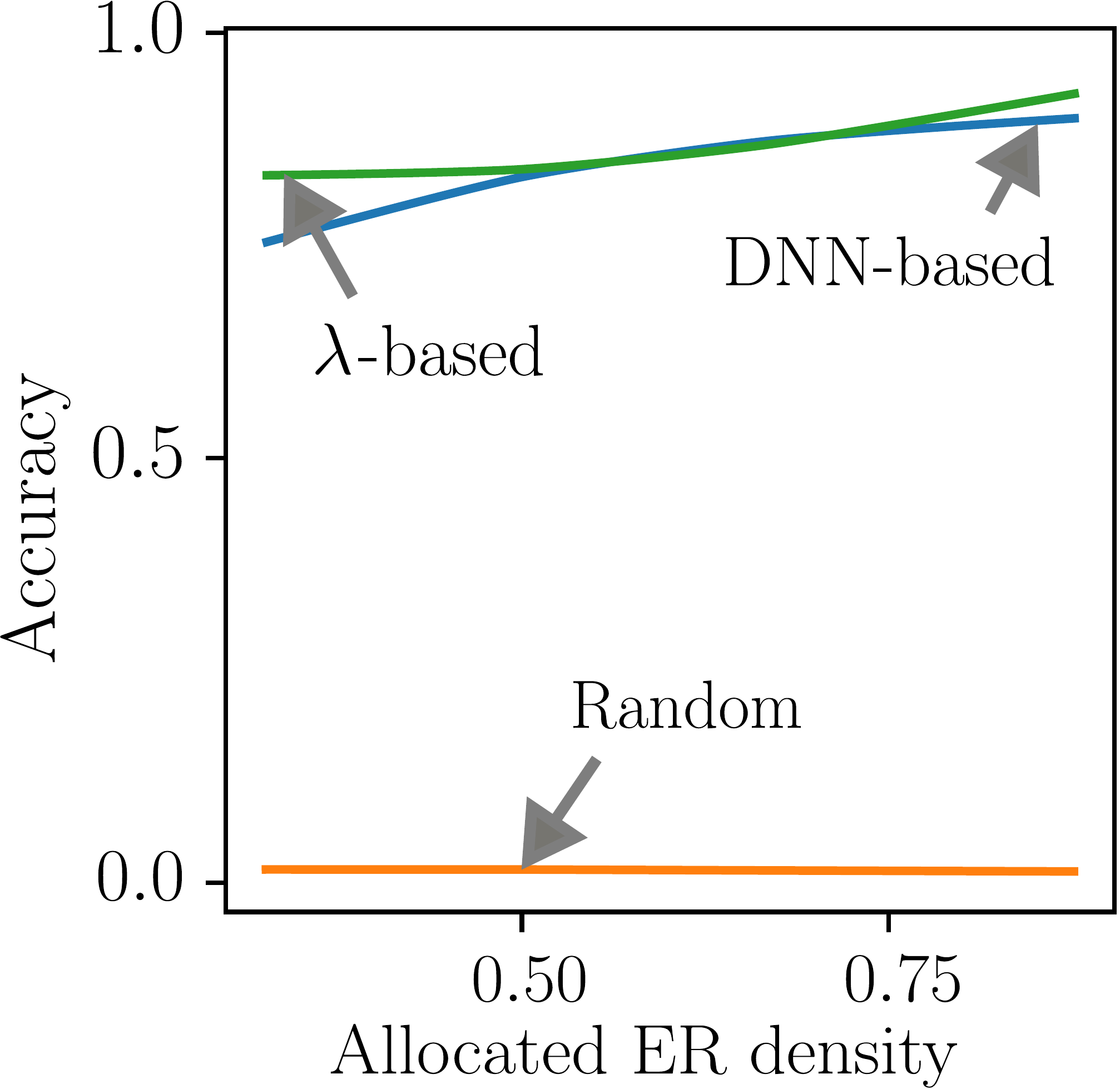}\label{fig:acc_DNNvsMu}}
\caption{Comparison of the DNN-based and $\mu$-based methods: (a) The mean square error; (b) The exact accuracy.}
\label{figs:DNNvsMu}
\end{figure}

\begin{figure}[h]
\centering
\includegraphics[width=0.55\linewidth]{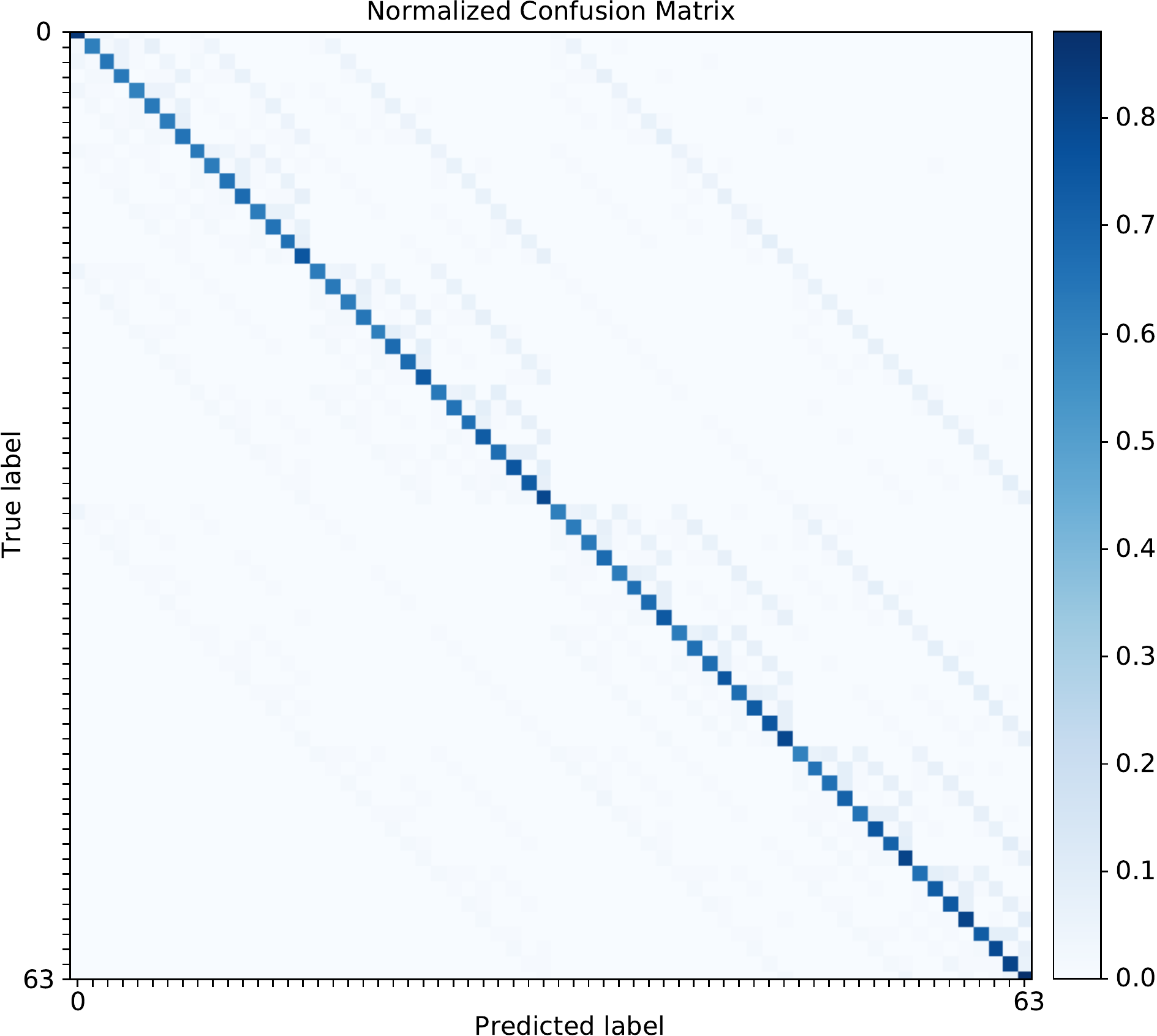}
\caption{The normalized confusion matrix of the trained DNN.}\label{fig:confusion}
\end{figure}

We first populate the training data by solving the optimization problem (\ref{eq:optim2}) offline, to find the optimal allocation vectors (as target labels) using the standard semidefinite problem solver tool CVX \cite{Grant2014CVX} for $194,000$ realizations of the network model with the aforementioned parameters. The proper set of hyper-parameters found are shown in Table \ref{tb:NNparams}. We dedicated $20\%$ of the data for testing purposes, and split the remaining $80\%$ into $80\%$ for the training and $20\%$ for the validation. While the training and validation sets are used in plotting Fig. \ref{fig:accuracy-mse}, the testing data is used in plotting Figs. \ref{figs:DNNvsLambda}-\ref{fig:confusion}.

As depicted in Fig. \ref{fig:NN1}, our DNN has three hidden layers, each of which is described in Table \ref{tb:NNparams}. The regularizer used in the first hidden layer is the $l1$-activity-regularizer with parameter $l1$ set to $0.001$. Activity regularizers allow us to apply penalties on layer activity during optimization. These penalties are incorporated in the binary cross-entropy loss function that the DNN tries to minimize during training. By doing this, we avoid over-fitting of the model. We used the well-known Adam optimizer \cite{Kingma2014Adam} with initial learning rate $0.01$ and a decay rate of $0.0027$. The batch size for training is set to $16$. The last layer has a sigmoid activation layer as explained before.

Figure \ref{fig:mse} shows the loss value of the training and validation datasets. As observed, the curves intersect at almost the last epoch, after which over-fitting will occur. Fig. \ref{fig:accuracy} shows the accuracy related to the training and validation datasets. The accuracy metric used here differs from the built-in binary accuracy metric in the {\it Keras} library which counts the fraction of all matches between each element of the predicted allocation vector with its corresponding true allocation vector. This is a customized metric to measure the fraction of \textit{exact} match between the target data and the predicted output data of the DNN. Fig. \ref{fig:accuracy} reveals that an accuracy of $76\%$ is reached with the proposed DNN for the SWIPT network. It should be noted that the AP can get paid even if there is no exact match between the predicted allocation vector and the true one. For example, let the true and predicted allocation vectors be $\mathbf{a}_{\rm true}=(1,1,1,0,0,0)^{T}$ and $\mathbf{a}_{\rm pred}=(1,1,0,0,0,0)^{T}$. Then, there happens an error in prediction. However, since the predicted allocation vector is a subset of the true allocation vector, the AP will get some profit from this fault prediction. It is also interesting to note that even if the predicted vector is not a subset of the true allocation vector or even if it is not a feasible allocation, the AP still has the chance to make profit. Recall that a subset $A^{(l)}$ is unfeasible if all the constraints of the subset cannot be covered with the AP's power budget $P$. However, it is still possible that the beamforming vectors found based on that subset do cover the constraints of a fraction of the devices. Therefore, the customized exact accuracy provides a lower-bound on the performance of the AP. It is interesting to mention that the achieved $76\%$ exact accuracy is equivalent to the built-in Keras binary accuracy of $95\%$. For the example mentioned above, the binary accuracy is $83.33\%$ but the exact accuracy is $0\%$.

Figure \ref{figs:DNNvsLambda} compares the accuracy and the mean square error (MSE) of the DNN-based solver with the proposed iterative $\lambda$-based approach for a WIT system with the same default system setting except that here $I=6$ and $J=0$. The curves are plotted in terms of the density of $1$'s in the label test dataset. As Fig. \ref{fig:acc_DNNvsLambda} shows, the accuracy of the $\lambda$-based method is comparable to the DNN-based method while the density of UEs in the system is either very low or very high. The low-density happens when the UEs' constraints cannot be well satisfied with the power budget of the AP; otherwise a high-density happens. In the middle-density range, the DNN-based outperforms the heuristic method.

Figure \ref{figs:DNNvsMu} compares the accuracy and the MSE of the DNN-based solver with the proposed iterative $\mu$-based approach for a WPT system with the same default system setting except that here $J=6$ and $I=0$. As observed, the accuracy of the heuristic method outperforms the DNN-based method particularly for extreme densities. Compared to the user density approach, the more exact way of demonstrating the accuracy of a network for a specific target label is by use of confusion matrices. These matrices are usable only for multi-class single-label classifications. However, by decoding our model's binary target labels to decimal-valued target labels, we can use the confusion matrix to evaluate the accuracy of the DNN for different classes. In fact, such decoded target labels are the same as subscript $l$ in $\mathbf{a}^{(l)}$. Fig.~\ref{fig:confusion} shows the normalized confusion matrix for the same default system setting. Thus, the decoded (One-Hot) targets will have $64$ labels. As can be inferred from the colors of the main diagonal of the confusion matrix, the normalized correct predictions range roughly from $70\%$ to $90\%$.

\begin{table}[h]
\centering
\caption{Accuracy of the DNN-Based Method for Different System Models}
\label{tb:AccuracyPerformance}
\begin{tabular}{c c c}
& Binary accuracy & Exact accuracy\\\toprule
SWIPT  & $95\%$ & $76\%$\\\hline
WIT  & $97\%$ & $80\%$\\\hline
WPT  & $97\%$ & $79\%$\\
\end{tabular}
\end{table}

Table \ref{tb:AccuracyPerformance} reveals the performance of the proposed DNN for three different user configurations with $K=6$ UEs: 1) the SWIPT system with $I=4$ and $J=2$; 2) the WIT system with $I=6$ IRs and no ERs; and the WPT system with $J=6$ ERs and no IRs.

\begin{table}[h]
\centering
\caption{Time Complexity of the Methods}\label{tb:NumericalComplexity}
\begin{tabular}{c c c c c}
& BnB-SDR & BnB-UDD & Heuristic-UDD & DNN-UDD\\\toprule
$K=4$  & $1.73$ & $0.11$ & $0.03$ & $0.04$\\\hline
$K=6$ & $10.55$ & $0.47$ & $0.11$ & $0.05$\\\hline
$K=8$ & $57.77$ & $3.13$ & $0.18$ & $0.06$
\end{tabular}
\end{table}

Table \ref{tb:NumericalComplexity} compares numerically the time complexity of the proposed methods with that of the BnB method for a WIT system. In BnB-SDR, each optimization problem related to each BnB branch is solved using the SDR technique \cite{Bengtsson1999BFSDR}, whereas in BnB-UDD, the UDD technique \cite{Schubert2004BeamformSINRconstraints} is applied to solve each BnB branch problem. As expected, since the UDD technique is much faster than the SDR technique, BnB-UDD is order of magnitudes less running-time complex than the BnB-SDR. We used our proposed BnB algorithm in Algorithm 1 (cf. Section \ref{sec:Algos}) for both BnB-SDR and BnB-UDD. The Heuristic-UDD and DNN-UDD methods are our proposed solutions. In Heuristic-UDD, the UDD method is used in each iteration of Algorithm 2. For $K=8$ devices in the system, $8$ runs of the UDD algorithm are required at worst-case, whereas in the BnB-UDD method $255$ runs of the UDD algorithm are required at worst-case. Hence, we see the huge difference between $0.18$ and $3.13$. In DNN-UDD, after finding the optimal set of devices by using the DNN-based approach\footnote{The same DNN architecture of Table \ref{tb:NNparams} is used, except that the input and output layers are modified according to the value of $K$}, the UDD algorithm is run once to output the optimal beamforming vector. The simulations are averaged over 1000 random realizations of the network setting. For small values of $K$, e.g., $K=4$, the heuristic method is a bit faster than the DNN-based method. For the larger value $K=8$, the DNN-based method runs three times faster than the heuristic method, $50$ times faster than BnB-UDD, and about $1000$ times faster than BnB-SDR.

Last but not least, it should be noted that while the DNN-based method outputs the solution in fewer clock cycles, the heuristic method has two attractive properties: i) it does not require training, and ii) it can easily scale to any network setting with arbitrarily number of devices.
\section{Conclusion}\label{sec:Conclusion}
We tackled the problem of finding the optimal revenue maximizing dominant-strategy incentive-compatible mechanism, namely, the allocation and payment rules, for a SWIPT network wherein a multi-antenna AP sells its spatially-multiplexed radio links to SINR-constrained information devices, and its power to energy harvesting devices. Having solved the MINLP revenue maximization problem by applying a proposed efficient Branch-and-Bound algorithm and by leveraging conventional optimization techniques, namely, semidefinite relaxation or uplink-downlink duality methods in each branch, we highlighted the time-greediness of such techniques for finding the optimal allocation rule, especially as the number of network devices increases. We also proposed an iterative suboptimal heuristic solution method with polynomial time complexity. Then, we designed and trained a DNN to find the allocation rule with linear time complexity and an accuracy of $76\%$.

\bibliographystyle{IEEEtran}
\bibliography{MyBibtexLib}
\end{document}